\documentclass[12pt]{article}
\usepackage{graphicx,psfrag,epsf}
\usepackage{enumitem}
\usepackage{url,enumerate, amssymb, amsfonts}
\usepackage[colorlinks = true,
linkcolor = blue,
urlcolor  = blue,
citecolor = blue,
anchorcolor = blue]{hyperref}
\usepackage{setspace,listings}
\usepackage{graphicx}
\usepackage{amsmath}
\usepackage{psfrag}
\usepackage[font=small,labelfont=bf]{caption}
\usepackage{enumerate}
\usepackage{authblk}
\usepackage{natbib}
\usepackage{url} 
\usepackage{setspace}
\usepackage{lscape}
\usepackage{color,amssymb}
\usepackage{mathtools}
\usepackage{dcolumn}
\usepackage{indentfirst, verbatim, float}
\usepackage{mathtools, amsthm, subcaption}
\theoremstyle{plain}
\newtheorem{theorem}{Theorem}
\theoremstyle{definition}

\newtheorem{corollary}{Corollary}
\newtheorem{lemma}{Lemma}
\newtheorem{proposition}{Proposition}
\usepackage{sidecap}
\usepackage{titlesec}
\usepackage{mathrsfs}
\usepackage[plain,noend, boxed, lined, algoruled]{algorithm2e}

\def\MGC{\textsc{mgc}}
\def\HHG{\textsc{hhg}}
\def\DMGC{\textsc{dmgc}}
\def\dCorr{\textsc{dcorr}}
\def\dCov{\textsc{dcov}}
\def\FH{\textsc{fh}}

\providecommand{\mc}[1]{\mathcal{#1}}

\usepackage{tikz}
\usetikzlibrary{shapes.geometric, arrows}
\tikzstyle{node} = [rectangle, rounded corners, minimum width=0.5cm, minimum height=0.5cm, draw=black, fill=white!30]
\tikzstyle{arrow} = [thick,->,>=stealth]

\sidecaptionvpos{figure}{c}

\def\spacingset#1{\renewcommand{\baselinestretch}%
{#1}\small\normalsize} \spacingset{1}

\title{\bf Network Dependence Testing via Diffusion Maps and Distance-Based Correlations}

\author{Youjin Lee}
\affil{Department of Biostatistics, Johns Hopkins School of Public Health}

\author{Cencheng Shen}
\affil{Department of Applied Economics and Statistics, University of Delaware}

\author{Carey E. Priebe}
\affil{Department of Applied Mathematics and Statistics, Johns Hopkins University}

\author{Joshua T. Vogelstein}
\affil{Department of Biomedical Engineering, Johns Hopkins University \thanks{jovo@jhu.edu}}
\date{}

\begin{document}

\maketitle

\begin{abstract}
Deciphering the associations between network connectivity and nodal attributes is one of the core problems in network science. The dependency structure and high-dimensionality of networks pose unique challenges to traditional dependency tests in terms of theoretical guarantees and empirical performance. We propose an  approach to test network dependence via diffusion maps and distance-based correlations. We prove that the new method yields a consistent test statistic under mild distributional assumptions on the graph structure, and demonstrate that it is able to efficiently identify the most informative graph embedding with respect to the diffusion time. The testing performance is illustrated on both simulated and real data. 
\end{abstract}

\noindent%

%%%%%%%%%%%%%%%%%%%%%%%%%%%
\sloppy
\spacingset{1.45}

\section{Introduction}
\label{sec:intro}

Network data has seen increased availability and influence in statistics, physics, computer science, biology, social science, etc., which poses many challenges due to its distinct structure. A network or graph is formally defined as an ordered pair $\mc{G}=(\mc{V},\mc{E})$, where $\mc{V}$ represents the set of nodes and $\mc{E}$ is the set of edges, and $n = |\mc{V}|$. The edge connectivity of a graph can be compactly represented by the adjacency matrix $\mc{A} = \{\mc{A}(i,j) : i,j= 1,..,n\}$, where $\mc{A}(i,j)$ is the edge weight between node $i$ and node $j$. For example, for an unweighted and undirected network, $\mc{A}(i,j) =\mc{A}(j,i) = 1$ if and only if node $i$ and node $j$ are connected by an edge, and zero otherwise.
Often, each node has some associated nodal attributes, which we denote as $X_i \in \mathbb{R}^{p}$ and use $\mc{X} = [X_1 | \cdots | X_n]$ to represent the collection of attributes.

This paper focuses on independence testing between network connectivity and nodal attributes. Assuming for the adjacency matrix $\mc{A}$ and attributes $\mc{X}$, the connectivity and attribute corresponding to each node are identically and jointly distributed as $F_{AX}$, the null and alternative hypotheses of interest are:
\begin{align}
\label{eq:hyp}
&H_{0}: F_{AX}=F_{A}F_{X}\\
&H_{A}: F_{AX} \neq F_{A}F_{X}. \nonumber
\end{align}
There are many network data examples where testing independence can be a crucial first step. For example, determining potential correlation between cultural tastes and relationships over social network~\citep{lewis2012social}, identifying association between the strength of functional connectivity and brain physiology such as regional cerebral blood flow in brain network~\citep{liang2013coupling}, embedding text data and its hyperlink networks jointly into a low-dimensional structure~\citep{ShenVogelsteinPriebe2016}. Sometimes the correlations among nodes are not proportional to the strength of connectivity between them. For instance, in signaling network of biological cells, reaction rate for each cell exhibits non-linear dependence on the neighboring response due to complex, cooperative biological process involved~\citep{hernandez2017nonlinear}. We can observe nonlinear dependence in concentrated propagration among a few focal persons in the social network~\citep{nekovee2007theory}, and in screening informative brain regions for sex and site difference from fMRI image graphs~\citep{mgc4}, etc. 

A notable obstacle in network inference is the structure of the edge connectivity. Namely, for an undirected graph, $\mc{A}$ is a symmetric binary matrix whose edges are not independent of each other, thus preventing many well-established methods from being directly applicable. 
One approach is to assume certain model on the graph structure, then solve the inference question based on the model assumption~\citep{wasserman1996logit, fosdick2015testing, howard2016understanding}. Another  approach is spectral embedding, which first embeds the $n \times n$ adjacency matrix $\mc{A}$ into an $n \times q$  matrix $\mc{U}$ by eigendecomposition, then directly works on $\mc{U}$~\citep{rohe2011spectral,SussmanEtAl2012, tang2017a}. For example, the network dependence test proposed by Fosdick and Hoff \citep{fosdick2015testing} assumes that the adjacency matrix is generated from a multivariate normal distribution of the latent factors, estimates the latent factor associated with each node from $\mc{A}$, followed by applying the standard likelihood ratio test on the normal distribution. 

However, model-based approaches are often limited by, and do not perform well beyond, the model assumptions. Moreover, spectral embedding is susceptible to misspecification of the dimension of $q$. Both of these factors can significantly degrade the later inference performance. Indeed, as a ground truth is unlikely in real networks \citep{Leto2017}, one often desires a method that is effectively non-parametric and robust against algorithm parameter selection~\citep{ChenShenVogelsteinPriebe2016}.

We propose a methodology to test network dependency via diffusion maps and distance-based  correlations, which is universally consistent under mild graph distributional assumptions and works well under many popular network models. The proposed method also overcomes parameter selection issues, and exhibits superior empirical testing performance. 
The \texttt{R} code and accompanying data are publicly available online at \url{http://neurodata.io/tools/mgc} and \url{https://github.com/neurodata/mgc}.

\section{Preliminaries}
\label{sec:pre}
\subsection{Notation}
\label{ssec:notation}

We denote a random variable by capital letter such as $X$ with distribution $F_{X}$, and denote a matrix or a set of vectors by calligraphic letter such as $\mc{X}$. 
For each node $i \in \mc{V}$, its attribute is denoted by $X_{i}$ whose realizations are in $\mathbb{R}^{p}$;
and its edge connectivity vector is denoted by $A_i \in \mathbb{R}^{n}$, which is a column in the $n \times n$ adjacency matrix $\mc{A}$.
We assume that $(X_i,A_i) \sim F_{XA}$, i.e., identically distributed attributes and connectivity vectors. 
Later we introduce a multiscale node-wise representation of the nodes as an $n \times q$ matrix $\mc{U}^{t} = [ U^{t}_{1} | U^{t}_{2} | \cdots | U^{t}_{n} ]$ for any $t \in \{0\} \bigcup \mathbb{Z}^{+}$, where $q$ is the embedding dimension and $t$ is the Markov iteration time step. 
Let $\cdot^{*}$ denote estimated optimality; $\cdot^{t}$ denotes either the $t^{\mbox{th}}$ power or  time step, which shall be clear in the context; and $\cdot^{T}$ is the matrix transpose.

\subsection{Diffusion Maps}
\label{ssec:method2}

Because the rows and columns of a symmetric adjacency matrix may be correlated, directly operating on the adjacency matrix breaks theoretical guarantees of existing dependence tests. 
The diffusion map is introduced as a feature extraction algorithm by Coifman and Lafon~\citep{coifman2005geometric,coifman2006diffusion,lafon2006diffusion}, which computes a family of embeddings in  Euclidean space by eigendecomposition on a diffusion operator of the given data. Here we introduce a version tailored to adjacency matrices.

To derive the diffusion maps for given observations of size $n$, the first step is to choose a $n \times n$ kernel matrix $\mc{K}$ that represents the similarity within the sample data. The adjacency matrix $\mc{A}$ is a natural similarity matrix; for undirected graphs we let $\mc{K}=\mc{A}$, for directed graphs  we let $\mc{K}= (\mc{A} + \mc{A}^{T})/2$.
The next step is to compute the normalized Laplacian matrix by 
\begin{align*}
\mc{L} = \mc{B}^{-1/2} \mc{K} \mc{B}^{-1/2},
\end{align*}
where $\mc{B}$ is the $n \times n$ degree matrix of $\mc{K}$. When $\mc{B}(i,i)$ or $\mc{B}(j,j)$ is zero, $\mc{L}(i,j)=0$.

The diffusion map $\mc{U}^{t}=\{U_{i}^{t} \in \mathbb{R}^{q} : i=1,\ldots,n\}$ is then computed by eigendecomposition, namely
\begin{align}
\label{eq:U}
U_{i}^{t}  &= \begin{pmatrix} \lambda^{t}_{1} \phi_{i1}, & \lambda^{t}_{2} \phi_{i2},   & \cdots, & \lambda^{t}_{q} \phi_{iq} \end{pmatrix}^{T} \in \mathbb{R}^{q}, \quad i = 1, \ldots, n,
\end{align}
where $\{ \lambda_{j}^t : j = 1,2,\ldots, q \}$ and $\{ \phi_{j} \in \mathbb{R}^{n} : (\phi_{1j}, \phi_{2j}, \ldots, \phi_{nj} ),~j=1,2,\ldots, q \}$ are the $q$ largest eigenvalues and corresponding eigenvectors of $\mc{L}$ respectively, and $\lambda^{t}_{j}$ is the $t^{\mbox{th}}$~power of the $j^{\mbox{th}}$~eigenvalue. 
The diffusion distance between the $i^{\mbox{th}}$~observation and the $j^{\mbox{th}}$~observation is defined as the weighted $\ell^{2}$ distance of the two points in the observation space, which equals the Euclidean distance in the diffusion coordinate:
\begin{align*}
%\label{eq:diffusion}
\mc{C}^{t}(i,j)  =   \| U_{i}^{t} - U_{j}^{t} \|, \quad i,j = 1,2, \ldots , n,
\end{align*}
where $\| \cdot \|$ is the Euclidean distance.

When $t=0$, the diffusion map is exactly the same as a normalized graph Laplacian embedding in \citet{rohe2011spectral} up-to a linear transformation; when $t>0$, the diffusion maps are weighted graph Laplacian by powered eigenvalues~\citep{lafon2006diffusion}; and the diffusion map at $t=1$ equals the adjacency spectral embedding up-to the degree constant~\citep{sussman2014consistent}. Therefore, the diffusion maps can be viewed as a single index family of  embeddings. The embedding dimension choice $q$ can be selected via the profile likelihood method in~\citet{ZhuGhodsi2006}, which is a standard algorithm in dimension reduction literature. To select the optimal $t$, we will utilize a smoothing technique to maximize the dependency, as discussed shortly.

%%%%%%%%%%%%%%%%%%%%%%%%%%%%%%%%%%%%%%%
\subsection{Distance-Based Correlations}
\label{ssec:method1}

The problem of testing general dependencies between two random variables has seen notable progress in recent years. The Pearson's correlation~\citep{Pearson1895} is the most classical approach, which determines the existence of linear relationship via a correlation coefficient in the range of $[-1,1]$, with $0$ indicating no linear association and $\pm 1$ indicating perfect linear association. To better capture the dependencies not limited to linear relationship, a variety of distance-based correlation measures have been suggested, including the distance correlation and energy statistic \citep{szekely2007measuring,szekelyRizzo2013a, RizzoSzekely2016}, kernel-based independence test \citep{GrettonGyorfi2010}, Heller-Heller-Gorfine test \citep{HellerGorfine2013,heller2016consistent}, and multiscale graph correlation~\citep{shen2017mgc,shen2016discovering},
among others. In particular, distance correlation is a distance-based dependency measure that is consistent against all possible dependencies with finite second moments. The kernel independence test is a kernel variant of distance correlation~\citep{sejdinovic2013equivalence,mgc5}. The multiscale graph correlation inherits the same consistency of distance correlation with better finite-sample testing powers under high-dimensional and nonlinear dependencies, via defining a family of local correlations and efficiently searching for the optimal local scale in testing. Here we briefly introduce distance correlation and multiscale graph correlation, which are denoted as $\dCorr$ and $\MGC$ in the equations.

Given $n$ pairs of sample data that are independently and identically distributed (i.i.d.), namely $(\mc{U}, \mc{X}) = \{  (U_{i}, X_{i} ) \stackrel{i.i.d.}{\sim} F_{UX} \in \mathbb{R}^{q} \times \mathbb{R}^{p}: i = 1,2, \ldots, n \}$. Denote the pairwise distances within $\{U_{i}\}_{i=1}^{n}$ and $\{X_{i}\}_{i=1}^{n}$ as $\mc{C}(i,j) = \| U_{i} - U_{j} \|$ and $\mc{D}(i,j) = \| X_{i} - X_{j} \|$ for $i,j=1,2, \ldots , n$ respectively. The sample distance covariance is denoted as 
\begin{align*}	 
%\label{eq:dCov}
\dCov_{n}(\mc{U},\mc{X}) = \frac{1}{n^2} \sum\limits_{i,j=1}^{n} \tilde{\mc{C}}(i,j) \tilde{\mc{D}}(i,j),
\end{align*}
where $\tilde{\mc{C}}=\mc{H} \mc{C} \mc{H}$ and $\tilde{\mc{D}}=\mc{H} \mc{D} \mc{H}$, and $\mc{H} =\mc{I}_{n \times n}- \mc{J}_{n \times n} / n$ is the centering matrix with $\mc{I}_{n \times n}$ being the $n \times n$ identity matrix and $\mc{J}_{n \times n}$ being the $n \times n$ matrix of all ones. The distance correlation follows by normalizing distance covariance via Cauchy-Schwarz into the range of $[-1,1]$. 
\citet{szekely2007measuring} shows that sample distance correlation converges to a population form, which is asymptotically $0$ if and only if independence,  resulting in a consistent statistic for testing independence. An unbiased sample version of distance correlation is later proposed to eliminate the sample bias in distance correlation~\citep{szekely2013distance, SzekelyRizzo2014}, which is the default implementation in this paper.

The multiscale graph correlation is an optimal local version of distance correlation that improves its finite-sample testing power. It first derives all local distance covariances as 
\begin{align*}
%\label{eq:MGC}
\dCov_{n}^{kl}(\mc{U},\mc{X}) = \frac{1}{n^2} \sum\limits_{i,j=1}^{n} \tilde{\mc{C}}^{k}(i,j) \tilde{\mc{D}}^{l}(i,j); \quad k = 1,\ldots, \kappa ,~l= 1, \ldots, \gamma,
\end{align*}
where $\kappa$ and $\gamma$ are the number of unique numerical values in $\mc{C}$ and $\mc{D}$ respectively; $\tilde{\mc{C}}^{k}(i,j) = \tilde{\mc{C}}(i,j) \mathbb{I}(R^{\mc{C}}_{ij} \leq k )$; $\mathbb{I}(\cdot)$ is the indicator function; and $R^{\mc{C}}_{ij}$ is a rank function of $U_{i}$ relative to $U_{j}$, i.e., $R^{\mc{C}}_{ij} =k$ if $U_{i}$ is the $k^{\mbox{th}}$ nearest neighbor of $U_{j}$, and define equivalently $\tilde{\mc{D}}^{l}(i,j) = \tilde{\mc{D}}(i,j) \mathbb{I}(R^{\mc{D}}_{ij} \leq l)$ for $\{X_i\}$. Then the local distance correlations $\{ \dCorr^{kl} \}$ are the normalizations of the local distance covariances into $[-1,1]$ via Cauchy-Schwarz. 
Among all possible neighborhood choices, the multiscale graph correlation equals the maximum local correlation within the largest connected component of significant local correlations, i.e.,
\begin{align*}
\MGC_{n}(\mc{U},\mc{X})=\dCorr_{n}^{(kl)^{*}}(\mc{U},\mc{X}), \mbox{ where } (kl)^{*}=\arg\max_{(kl)}\mc{S}(\dCorr_{n}^{kl})
\end{align*}
for a smoothing operation $\mc{S}(\cdot)$ that filters out all in-significant local correlations. The multiscale graph correlation has been shown to have power almost equal or better than distance correlation throughout a wide variety of common dependencies, while being computationally efficient~\citep{shen2017mgc}. 

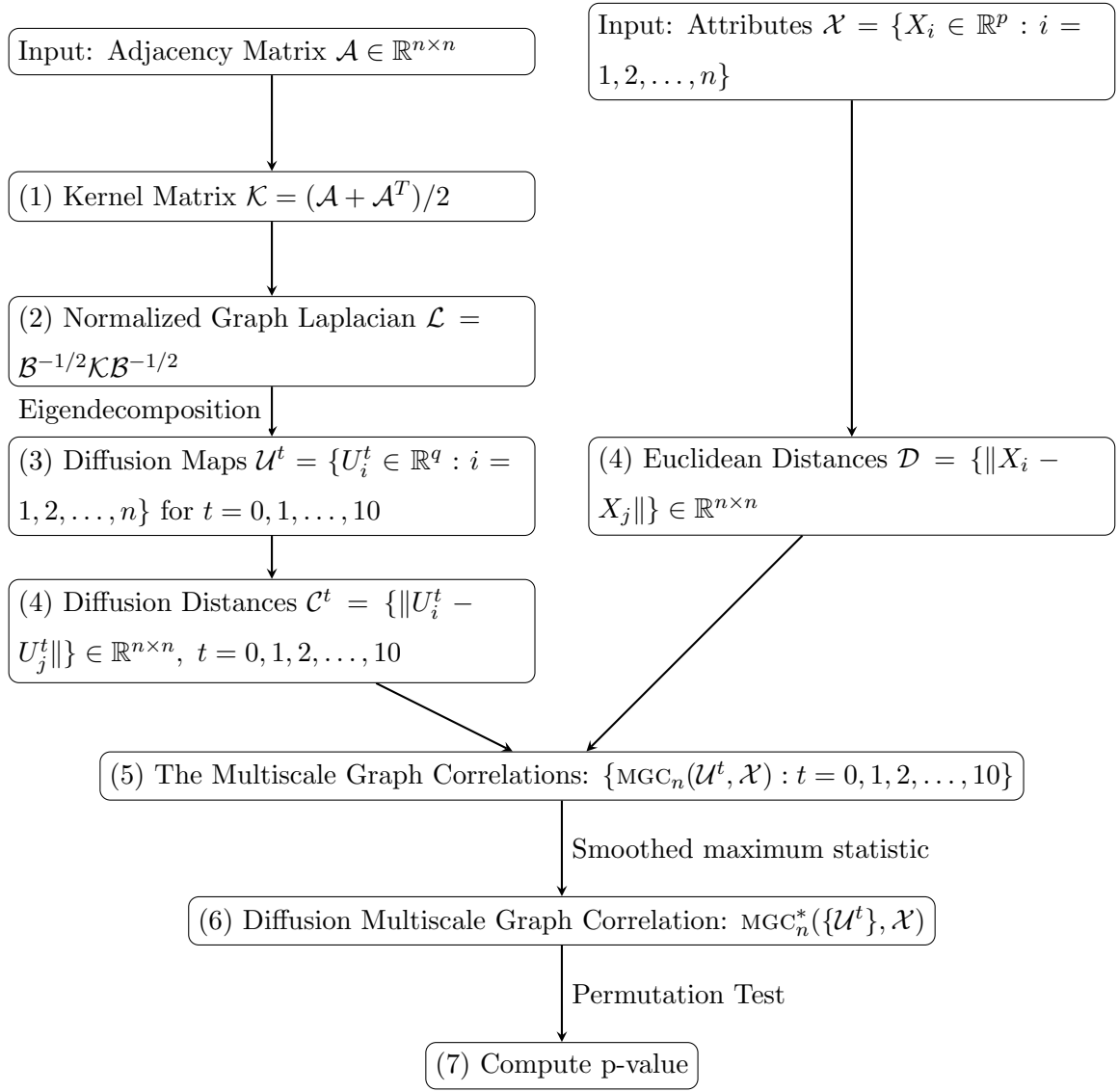
\begin{figure}[H]
	\small	
	\begin{tikzpicture}[node distance=2.0cm, every node/.style={rectangle,rounded corners, minimum width=0.5cm, minimum height=0.5cm, draw=black, fill=white!30, draw}]
	%\tikzalign{node} = [rectangle, rounded corners, minimum width=0.5cm, minimum height=0.5cm, draw=black, fill=white!30]
	%\tikzalign{arrow} = [thick,->,>=stealth]
	\node (in1) [xshift=1cm, text width=7cm] {Input: Adjacency Matrix $\mc{A} \in \mathbb{R}^{n \times n}$};
	\node (in2) [xshift=9cm, text width=7cm] {Input: Attributes $\mc{X}=\{X_{i} \in \mathbb{R}^{p} : i=1,2,\ldots,n \}$ };
	\node (S1) [below of=in1, text width=7cm] {(1) Kernel Matrix $\mc{K}=(\mc{A}+\mc{A}^{T})/2$};
	\node (S2) [below of=S1, text width=7cm] {(2) Normalized Graph Laplacian   $\mc{L} = \mc{B}^{-1/2} \mc{K} \mc{B}^{-1/2}$};
	\node (S2d) [below of=S2, text width=7cm] {(3) Diffusion Maps  $\mc{U}^{t} = \{U^{t}_{i} \in \mathbb{R}^{q} : i=1,2,\ldots, n \}$   for $t = 0,1, \ldots, 10$};
	
	\node (S3) [below of=S2d, text width=7cm] {(4) Diffusion Distances $\mc{C}^{t} =  \{\| U^{t}_{i}- U^{t}_{j}\|\} \in \mathbb{R}^{n \times n},~t = 0,1,2, \ldots, 10$};
	\node (S2x) [below of=in2, yshift=-4cm, text width=7cm] {(4) Euclidean Distances $\mc{D}= \{\| X_{i}- X_{j}\|\} \in \mathbb{R}^{n \times n}$};
	
	\node (D1) [below of=S3, xshift=4cm] {(5) The Multiscale Graph Correlations: $\{ \MGC_{n}(\mc{U}^{t},\mc{X}) : t = 0,1,2, \ldots, 10 \}$};
	\node (D2) [below of=D1] {(6) Diffusion Multiscale Graph Correlation: $\MGC_{n}^{*}(\{\mc{U}^{t}\}, \mc{X})$};
	\node (D3) [below of=D2] { (7) Compute p-value};
	
	\draw [thick,->,>=stealth, align=right] (in1) -- node[anchor=east, draw = none] {}   (S1);
	\draw [thick,->,>=stealth] (in2) -- (S2x);
	\draw [thick,->,>=stealth] (S2) -- node[anchor=east, draw = none] {Eigendecomposition} (S2d);
	%\draw [arrow] (S2) -- node;
	\draw [thick,->,>=stealth] (S1) -- (S2);
	\draw [thick,->,>=stealth] (S2d) -- (S3);
	\draw [thick,->,>=stealth] (S3) -- (D1);
	\draw [thick,->,>=stealth] (S2x) -- (D1);
	\draw [thick,->,>=stealth] (D1) -- node[anchor = west, draw=none]{Smoothed maximum statistic}(D2);
	\draw [thick,->,>=stealth] (D2) -- node[anchor=west, draw = none] {Permutation Test}(D3);
	\end{tikzpicture}
	\caption{Flowchart for Network Dependence Testing via Diffusion Multiscale Graph Correlation (\DMGC).}
	\label{fig:flow}
\end{figure}

\section{Main Results}
\label{sec:method}

\subsection{Testing procedure of Diffusion Correlation}
\label{ssec:method}

\begin{algorithm}[!ht]
	\caption{Testing procedure of Diffusion Correlation}
	\label{alg:flow}
	\enspace	Input: Adjacency matrix $\mc{A} \in \mathbb{R}^{n \times n}$ and nodal attributes $\mc{X} = \{ X_{i} \in\mathbb{R}^{p}: i = 1,2,\ldots,n \}$. \\
	\quad (1) Symmetrize $\mc{A}$ by $\mc{K} = (\mc{A} + \mc{A}^{T})/2$.\\
	\quad (2) Obtain normalized graph Laplacian matrix $\mc{L} = \mc{B}^{-1/2} \mc{K} \mc{B}^{-1/2}$. \\ \quad (3)  Do eigendecomposition to obtain diffusion maps $\mc{U}^{t} = \{ U^{t}_{1}, U^{t}_{2}, \ldots, U^{t}_{n} \}$ for $t=0,1,2,\ldots, 10$. \\
	\quad (4) Derive $n \times n$ Euclidean distance of diffusion map $\mc{C}^{t}$, i.e., diffusion distance, across $t$, and $n \times n$ Euclidean distance of nodal attributes, $\mc{D}$. \\
	\quad  (5) Compute the multiscale graph correlations using two distance matrices, $\mc{C}^{t}$ and $\mc{D}$, for~$t=0,1,\ldots, 10$. \\
	\quad  (6) Derive the diffusion multiscale graph correlation: $\MGC^{*}_{n} \left(\{\mc{U}^{t}\}, \mc{X} \right)$ by estimating $t^*$. \\
	\quad (7) Compute p-value using permutation test.\\
	\enspace Output: P-value at the estimated optimal step $t^{*}$, the estimated optimal time step $t^{*}$, 
	%\quad \quad
	dimension choice of $q$ via profile likelihood method, 
	%\quad \quad 
	multiscale local correlation maps $\{ \dCorr_{n}^{kl}(\mc{U}^{t}, \mc{X}) \}$, the optimal neighborhood choice $(k^{*}, l^{*})$.  
\end{algorithm}
Here we develop diffusion multiscale graph correlation, which synthesizes diffusion map embedding, multiscale graph correlation, and smoothed maximum to better test network dependency. A flowchart of the testing procedure is illustrated in Fig.~\ref{fig:flow}, and the details of each step are described in Algorithm~\ref{alg:flow}. 

The algorithm is flexible in the choice of correlation measures: by following the exact same steps but replacing the multiscale graph correlation by distance correlation in Step (5), one can compute the diffusion distance correlation. Similarly one can derive the diffusion Heller-Heller-Gorfine method. 
The motivation of the smoothing Step (6) is the following: suppose there exists an optimal $t$ for detecting the relationship between edge connectivity and attributes, then the test statistics at adjacent time steps $t-1$ and $t+1$ should also exhibit strong signal. 
Under independence, a large test statistic at certain $t$ can occur by chance and cause a direct maximum to have a low testing power, while the smoothed maximum effectively filters out any noisy and isolated large test statistic. In practice, it suffices to consider $t \in [0,1,\ldots,10]$ or even smaller upper bound like $3$ or $5$. When smoothed maximum does not exist, we set $t=3$ as the default choice.
The permutation test in Step (7) is a common nonparametric procedure used for real data testing in almost all dependency measures, which is valid as long as the observations are exchangeable under the null \citep{RizzoSzekely2016}.

\subsection{Theoretical Properties Under Exchangeable Graph}
\label{ssec:theory}
To derive the theoretical consistency of our methodology, the following mild assumptions are required on the distribution of the graph and the nodal attributes. \\

\indent (C1)  Graph $\mc{G}$ is an induced subgraph of an infinitely exchangeable graph. Namely, the adjacency matrix $\mc{A}$ satisfies
\begin{align}
\mc{A}(i,j) \stackrel{d}{=} \mc{A}(\sigma(i),\sigma(j))
\end{align}
for any $i,j=1,2,\ldots,n$ and any permutation $\sigma$ of size $n \in \mathbb{N}$. The notation $\stackrel{d}{=}$ stands for equality in distribution. \\
\indent (C2) Each nodal attribute $ X_i$ is generated independently and identically from $F_{X}$ with finite second moment. \\
\indent (C3) The matrix $\mc{A}$ is constrained to a domain $\Omega$ where the diffusion map embedding from $\mc{A} \in \Omega$ to $\mc{U}^{t}$ is injective for some $t$. \\
%\end{description}

Condition (C1) states that $\mc{G}$ is a collection of independently sampled nodes and their induced subgraph~\citep{orbanz2015bayesian, tang2017a, orbanz2017subsampling}, which is a distributional assumption satisfied by many popular statistical networks models. Based on condition (C1), the diffusion map $\mc{U}^{t}$ at each $t$ can furnish exchangeable and asymptotic conditional i.i.d. embedding for the set of nodes $\mc{V}(\mc{G})$.

\begin{theorem}
	\label{theorem:iid}
	Assume $\mc{G}$ satisfies (C1). Then at each fixed $t$, the embedded diffusion maps $\mc{U}^{t} = \{U_{i}^{t},~i=1,2,\ldots,n\}$ by Equation~\ref{eq:U} are exchangeable. As a result, there exists an underlying variable $\theta^{t}$ distributed as the limiting empirical distribution of ~$\mc{U}^{t}$, such that $U_{i}^{t} \mid \theta^{t} $ are asymptotically independently and identically distributed for $i=1,2, \ldots,n$ as $n \rightarrow \infty$.  
\end{theorem} 
Due to condition (C1), the permutation test is applicable to any $\mc{U}^{t}$ from an exchangeable sequence.
Condition (C2) is merely a regularity condition, and the distribution of $U_{i}^{t}$ automatically satisfies the same finite-moment assumption as shown in the Supplementary Material for proof of Theorem~\ref{theorem:convergence}. 
We then have consistency between the diffusion map at each $t$ and the nodal attribute. 

\begin{theorem}
	\label{theorem:convergence}
	Assume the graph $\mc{G}$ and the nodal attributes satisfy condition (C1) and (C2). Then as $n \rightarrow \infty$, the multiscale graph correlation between the diffusion map $\mc{U}^{t}$ at any fixed $t$ and the nodal attributes $\mc{X}$ satisfies:
	\begin{align*}
	\MGC_{n}(\mc{U}^{t},\mc{X}) \rightarrow c \geq 0,
	\end{align*}
	with equality if and only if $F_{U^{t} X} =F_{U^{t}} F_{X}$.
\end{theorem} 

The testing consistency naturally extends to the diffusion correlation, which further holds between edge connectivity and nodal attributes if condition (C3) is satisfied.
\begin{theorem}
	\label{theorem:time}
	Under the same assumption in Theorem~\ref{theorem:convergence}, it holds that
	\begin{align*}
	\MGC_{n}^{*}(\{\mc{U}^{t}\}, \mc{X}) \rightarrow c \geq 0,
	\end{align*}
	with equality if and only if $F_{U^{t} X} =F_{U^{t}} F_{X}$ for all $t \in [0,10]$. Therefore, the diffusion multiscale graph correlation is a valid and consistent statistic for testing independence between the diffusion maps $\{\mc{U}^{t}\}$ and nodal attributes $\mc{X}$. 
	
	If condition (C3) holds, then $\MGC^{*}_{n}(\{\mc{U}^{t}\}, \mc{X})$ is also valid and consistent for testing independence between the adjacency matrix and nodal attributes, i.e., it converges to $0$ if and only if the nodal attribute $X$ is independent of the node connectivity $A$.
\end{theorem}

\begin{corollary}
	\label{corollary:main}
	Theorem~\ref{theorem:time} still holds, when any of the following changes are applied to the testing procedure described in Section~\ref{ssec:method}: \\
	%\indent (1) The nodal attributes in the input are replaced by a second graph of the same node set, followed by the same diffusion map process; \\
	\indent (1) The multisclale graph correlation in step 2 is replaced by distance correlation or the Heller-Heller-Gorfine statistic; \\
	\indent (2) When $\mc{A}$ is restricted to be symmetric, binary, and of finite rank $q <n$, then condition (C3) holds at $t=1$.
\end{corollary}

Namely, point (1) suggests that under diffusion maps, other consistent dependency measure can also be used to produce a valid and consistent diffusion correlation, which enables us to compare a number of diffusion correlations in the simulations. Point (2) offers an example of random matrix $\mc{A}$ where the diffusion map is guaranteed injective within the domain.

\subsection{Consistency Under Random Dot Product Graph}
\label{ssec:theory2}
In this section, we illustrate the theoretical results via the random dot product graph model, which is widely used in network modeling. It assumes that each node has a latent position $W_{i} \overset{i.i.d.}\sim F_{W}$ for $i=1,2,\ldots, n$, and the edge probability $pr(\mc{A}(i,j) = 1 \mid W_{i}, W_{j})$ is determined by the dot product of the latent positions, i.e., 
\begin{align*}
\mc{A}(i,j)  \mid W_{i}, W_{j} \overset{i.i.d.}{\sim} Bernoulli\big( \langle W_{i},W_{j} \rangle \big),~i,j=1,2,\ldots, n \mbox{ and } i<j,
\end{align*}
under the restriction that all $W_{i}$'s are non-negative vectors and the dot product must be normalized within $[0,1]$. 

A random dot product graph is an exchangeable graph model that satisfies condition (C1). In addition, random dot product graph fully specifies all exchangeable graph models that are unweighted and symmetric, whose probability generating matrix $\mc{P}(i,j)=\langle W_{i},W_{j}\rangle$ is positive semi-definite.
\begin{proposition}[\citet{sussman2014consistent}] 
	\label{prop1}
	An exchangeable random graph has a finite rank $q$ and positive semi-definite link matrix $\mc{P}$, if and only if the random graph is distributed according to a random dot product graph with i.i.d. latent vectors $\{W_{i} \in \mathbb{R}^{q},i=1,\ldots,n\}$. 
\end{proposition}

Indeed, many other popular network modelings are special cases of random dot product graph, including the stochastic block model~\citep{airoldi2008mixed, hanneke2009network, rohe2011spectral, xin2017continuous}, its degree-corrected version~\citep{karrer2011stochastic}, the latent factor model from \citet{fosdick2015testing}, etc. 

\begin{proposition}[\citet{rohe2011spectral}]
	\label{prop2}
	Let $\mc{L}$ be the normalized graph Laplacian for an adjacency matrix $\mc{A}$ generated by a random dot product graph with latent positions of which construct the matrix of $\mc{W}=[W_{1}|W_{2}|\ldots|W_{n}] \in \mathbb{R}^{q \times n}$. Let~$\mc{U}^{t=1} =  [ U^{t=1}_{1}| U^{t=1}_{2}| \ldots| U^{t=1}_{n} ] \in \mathbb{R}^{q \times n}$. 
	Then there exists a fixed diagonal matrix $\mc{M}$ and an orthonormal rotational matrix $\mc{Q} \in \mathbb{R}^{q \times q}$ such that $\| \mc{U}^{t=1} - \mc{Q} \mc{M} \mc{W} \| \rightarrow 0$ almost surely.
\end{proposition}

Therefore, under random dot product graph, the diffusion map $\mc{U}^{t=1}$ asymptotically equals the latent position $\mc{W}$ up to a linear transformation. As the latent position under random dot product graph can be asymptotically recovered by diffusion maps, diffusion correlation is consistent against testing general dependency between $\mc{A}$ and $\mc{X}$ under random dot product graph.

\begin{corollary}
	\label{thm:AvsX}
	Under an induced subgraph from exchangeable graph with positive semi-definite link function, the diffusion multiscale graph correlation is consistent for testing independence between edge connectivity and nodal attributes. 
\end{corollary}

\subsection{Discussion on the Conditions}
Here we discuss the robustness of the methodology with respect to condition (C1)-(C3), and what happens when any of them is violated. These conditions are essential to guarantee a consistent and valid testing framework in general, which are not just limited to network dependence testing.

Condition (C1) is a crucial condition for the permutation test to be valid. When it is violated and neither set of data can be assumed exchangeable, all aforementioned test statistics may no longer be valid because the permutation test fails to control the type 1 error level as demonstrated in~\cite{Mantel2013}. In certain special cases like testing independence between two stationary times series, block permutation technique can be used to yield a valid test \citep{lacal2018estimating}, which can be readily used here but is not guaranteed valid for general non-exchangeable data.
Condition (C2) is a regularity condition required for distance-based correlation measure to be well-behaved, without which the distance variance can explode to infinity and cause the correlation measure to be ill-behaved.

In comparison, the diffusion correlation methodology is still valid without condition (C3). However, the second part of Theorem~\ref{theorem:time} will no longer hold, and the methodology is no longer universally consistent. Namely, certain signals of dependency may be lost during the diffusion map embedding procedure. As a result, the diffusion correlation could be asymptotically $0$ for some dependencies and thus no longer able to detect all possible dependencies between the edge connectivity and nodal attributes. In the Supplementary Material we illustrate the performance of the test statistics under the violation of positive semi-definite link function, and show relative robustness of distance-based tests compared to model-based tests when condition (C3) is violated. 

\section{Numerical Studies}
\label{sec:simulation}

\subsection{Stochastic Block Model}

Throughout the numerical studies, we compare diffusion multiscale graph correlation, diffusion distance correlation, diffusion Heller-Heller-Gorfine method, the Fosdick-Hoff likelihood ratio test~\citep{fosdick2015testing}, and direct embedding-based tests: using the adjacency spectral embedding and the latent factors to embed the adjacency matrix first, followed by any of the multiscale graph correlation, the distance correlation, or the Heller-Heller-Gorfine method. 
For each simulation, we generate a sample graph and the corresponding attributes, compute the test statistic of each method, carry out the permutation test with $r=500$ random permutations, and reject the null if the resulting p-value is less than $\alpha = 0.05$. The testing power of each method equals the percentage of correct rejection out of $m = 100$ replicates, and a higher power implies a better method against the given dependency structure.

The first simulation samples graphs from the stochastic block model. It assumes that each of the $n$ nodes in $\mc{G}$ must belong to one of $K \in \mathbb{N}$ blocks, and determines the edge probability based on the block-membership of the connecting nodes: For $i=1,\ldots,n$, assume there exists a latent variable of $Z_{i} \overset{i.i.d.}{\sim} Multinomial\big( \pi_{1}, \pi_{2}, ... , \pi_{K} \big)$ denoting the block-membership of each node, and denote the edge probability between any two nodes of class $k$ and $l$ as $b_{kl} \in \{0,1\}$. Then the upper triangular entries of $\mc{A}$ are independently and identically distributed when conditioning on $\mc{Z} = \{Z_{i}:~i=1,2,\ldots, n \}$:
\begin{align*} 
\mc{A}(i,j) \mid Z_{i}, Z_{j} \overset{i.i.d.}{\sim} Bernoulli\big\{ \sum\limits_{k,l=1}^{K} b_{kl} \mathbb{I} \big( Z_{i} = k, Z_{j} = l  \big)  \big\}; \quad  i < j,~i,j = 1,2, \ldots, n,
\end{align*}
where $\mathbb{I}(\cdot)$ is the indicator function. The sample data is generated at $n=100$ by using following parameters:
\begin{equation}
\begin{split}
\label{eq:Three}
& Z_{i} \overset{i.i.d.}{\sim} Multinom(1/3, 1/3, 1/3),\\
& \mc{A}(i,j) \mid Z_{i}, Z_{j} \sim Bernoulli \left\{0.5 \mathbb{I}(|Z_{i} - Z_{j}| = 0) + 0.2 \mathbb{I}(|Z_{i} - Z_{j}| = 1) + 0.4 \mathbb{I}(|Z_{i} - Z_{j}| = 2) \right\}, \\
&X_{i}\mid Z_{i}  \sim Multinom[ \{1+\mathbb{I}(Z_{i}=1)\}/4,~\{1+\mathbb{I}(Z_{i}=2)\}/4,~\{1+\mathbb{I}(Z_{i}=3)\}/4 ],
\end{split}
\end{equation} 
where $\mc{X}$ is a randomly polluted block assignment: for each $i$, $X_{i}=Z_{i}$ with probability $0.5$, and equally likely to take other values in $\Omega$, i.e., the true block-membership is observed half of the time. For the adjacency matrix, the within-block edge probability is always $0.5$; while the between-block edge probability is $0.2$ when the block labels differ by $1$, and $0.4$ when the block labels differ by $2$. As the edge probability between a node of block $1$ and a node of block $3$ is higher than the edge probability between block $1$ and block $2$, this three-block stochastic block model generates a noisy and nonlinear dependency structure between $\mc{A}$ and $\mc{X}$, and we would like to verify how successful the methods are in detecting the dependency between the adjacency matrix $\mc{A}$ and the noisy block assignment $\mc{X}$.

\begin{figure}[H]
	\centering
	\includegraphics[width=0.4\textwidth]{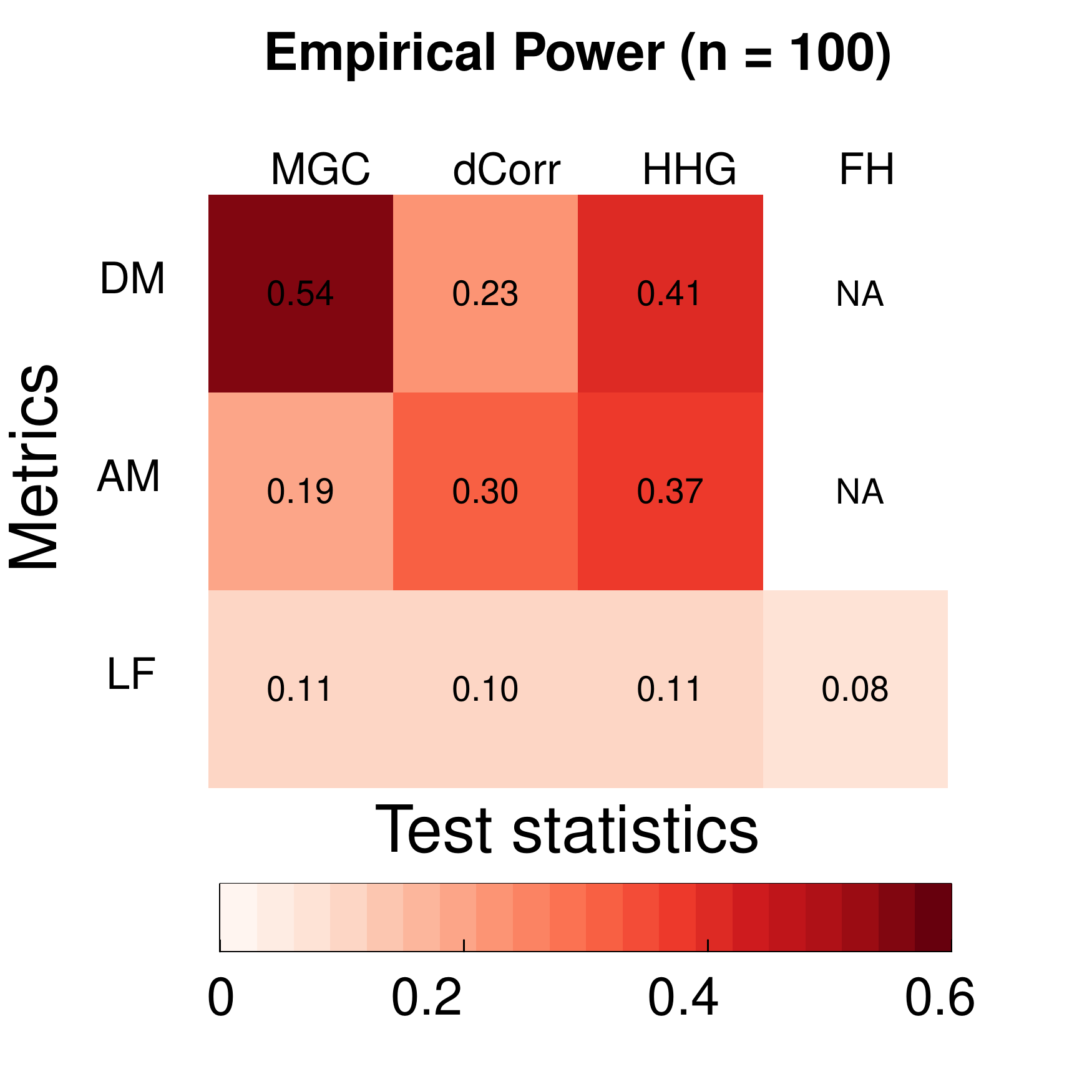}
	\caption{\label{fig:threeSBM} The testing powers for the three-block stochastic block model in Equation~\ref{eq:Three}. The y-axis lists the embedding choices: diffusion map (DM), adjacency spectral embedding (AM), and latent factor embedding (LF). The x-axis corresponds to the correlation measure in use: the multiscale graph corelation (MGC), distance correlation (dCorr), Heller-Heller-Gorfine (HHG), and the Fosdick-Hoff method (FH). The top three entries in the first row represent the diffusion correlation methods proposed in this paper, which outperform other embedding choices with diffusion multiscale graph correlation having the best power. }	
\end{figure}

Figure~\ref{fig:threeSBM} shows that diffusion multiscale graph correlation prevails the testing powers among all the methods, because multiscale graph correlation captures high-dimensional nonlinear dependencies better than distance correlation and Heller-Heller-Gorfine. A visualization of the sample data is available in Fig.~\ref{fig:embedding}(a).

\subsection{Stochastic Block Model with Linear and Nonlinear Dependencies}

To further understand and demonstrate the advantage of the diffusion approach, here we use the same three-block stochastic block model and its block-membership $\{ Z_{i} : i=1,2, \ldots, n=100 \}$ as in the previous section, except that the edge probability is now controlled by $\beta \in (0, 1)$ for all $i,j = 1, \ldots, n$:
\begin{equation}
\mc{A}(i,j) \mid Z_{i}, Z_{j} \sim Bernoulli \left\{ 0.5 \mathbb{I}(|Z_{i} - Z_{j}| = 0) + 0.2 \mathbb{I}(|Z_{i} - Z_{j}| = 1) + \beta \mathbb{I}(|Z_{i} - Z_{j}| = 2) \right\}.
\label{eq:mono}
\end{equation}
The noisy block-membership $\mc{X}$ is generated in the same way as before.
When $\beta = 0.2$, the three-block stochastic block model is the same as a two-block stochastic block model, where within-block edge probability equals $0.5$ while the between-block edge probability is always $0.2$, i.e., it represents a linear association between the adjacency matrix and the block-membership. When $\beta <0.2$, the dependency is still monotonic. When $\beta > 0.2$ and gets further away, the relationship becomes strongly nonlinear.
\begin{figure}[ht]
	\centering	
	\begin{minipage}[b]{0.49\textwidth}
		\includegraphics[width=\textwidth]{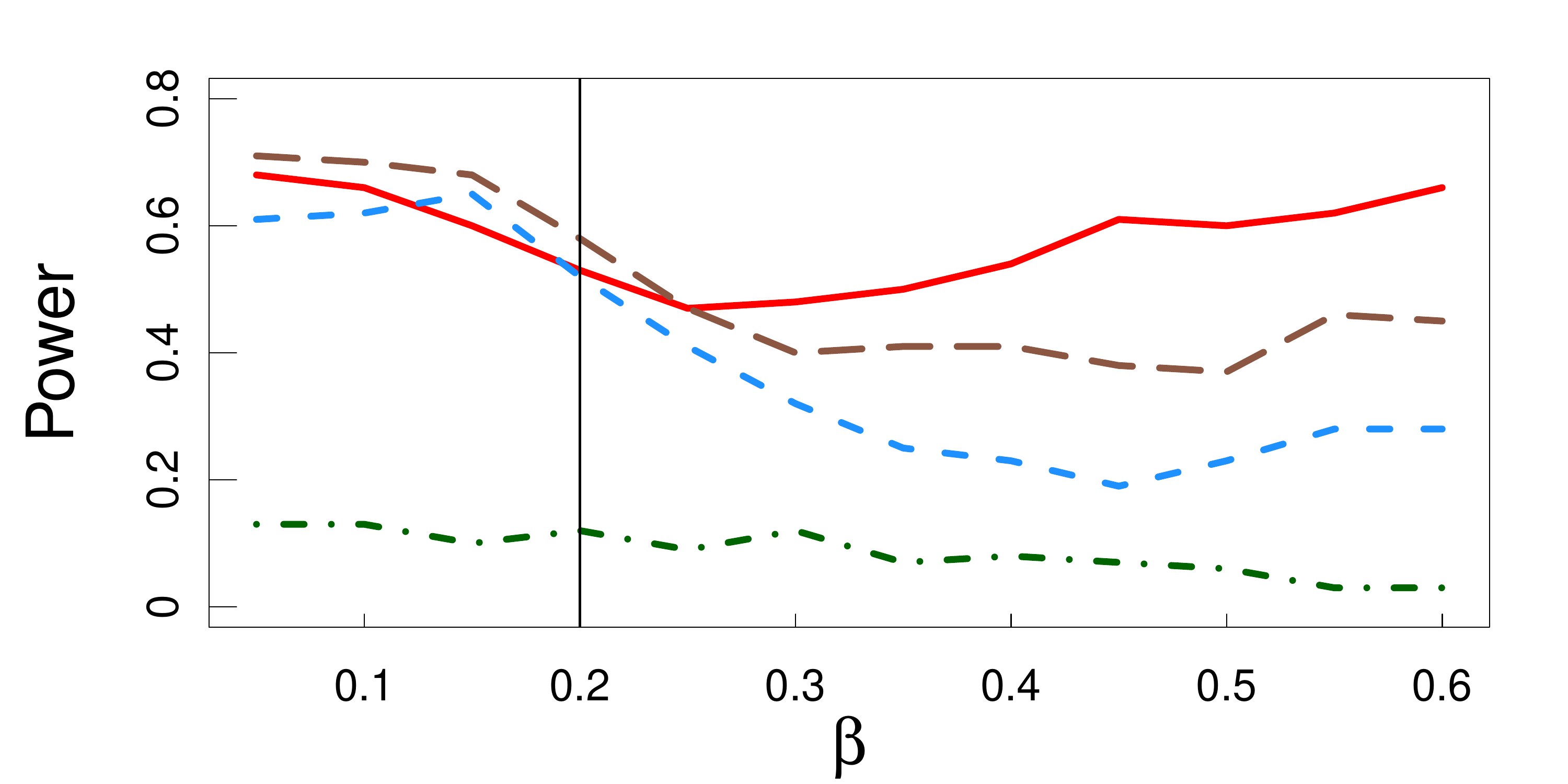} \\
		\centering (a)
	\end{minipage}
	\begin{minipage}[b]{0.49\textwidth}
		\includegraphics[width=\textwidth]{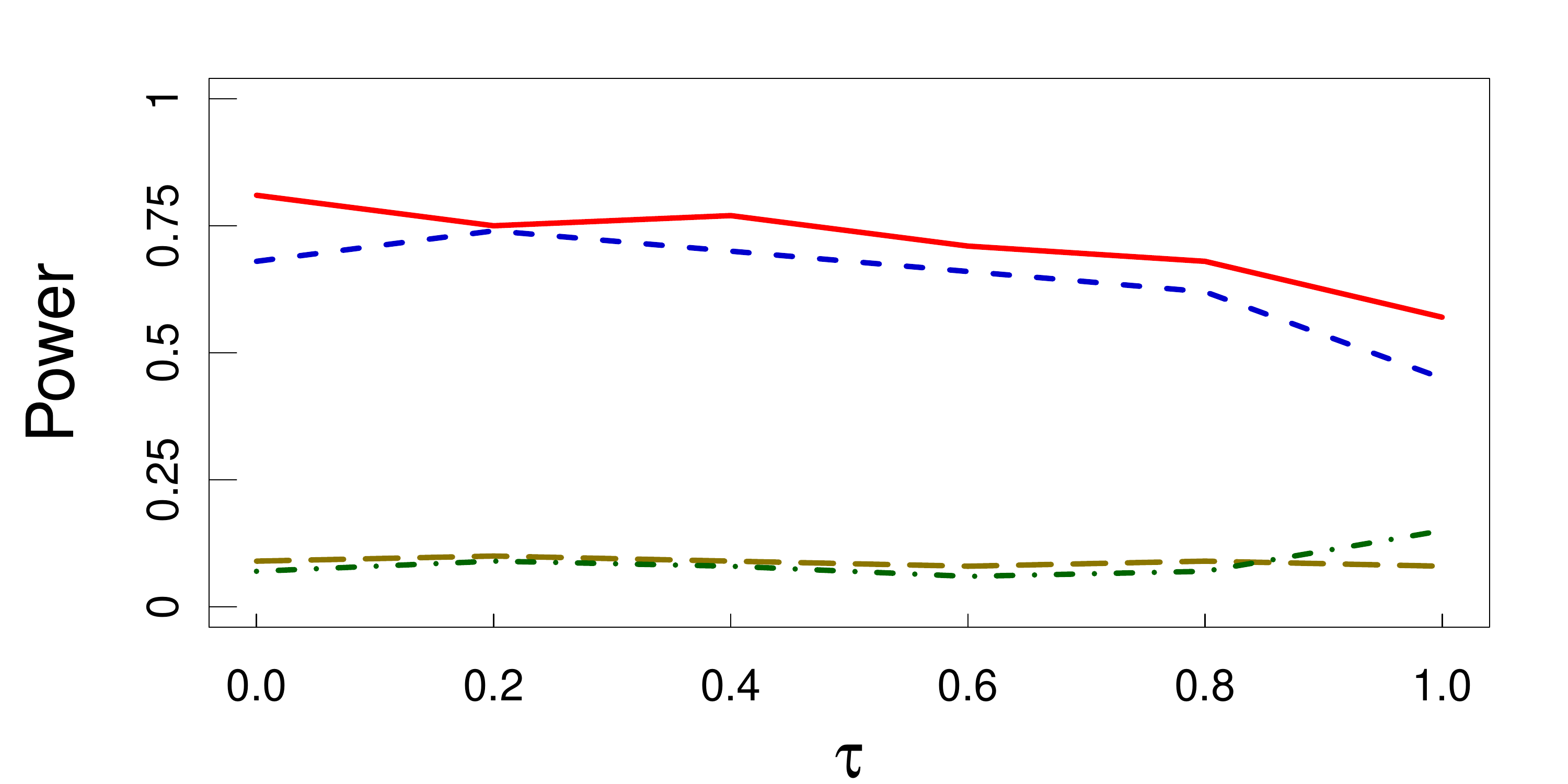}\\
		\centering (b)
	\end{minipage}
	\caption{(a) The power curve with respect to increasing $\beta$ under three-block stochastic block model (Equation~\ref{eq:mono}). When $\beta$ shifts from less than $0.2$ to higher than $0.2$, it represents a structural change in the relationship from monotone to non-monotone. Among all methods utilizing diffusion maps, diffusion \MGC~(solid red line) is evidently the best performing one comparing to diffusion \dCorr~(yellow brown dashes), diffusion \HHG~(blue small dashes), and \FH~(green dot-dash) test. (b) The power curve with respect to increasing $\tau$ under degree-corrected stochastic block model (Equation~\ref{eq:tau}). The edge variability increases as $\tau$ increases. Diffusion \MGC~(red solid) is relatively stable in power against increasing variability. The adjacency spectral embedding followed by \MGC~(dark blue dashes) is slightly worse, while the latent factor embedding followed by \MGC~(light yellow dashes) and \FH~(green dot-dash) have almost no power against all levels of $\tau$.}
	\label{fig:powerplot}
\end{figure}
Figure~\ref{fig:powerplot}(a) plots the power against $\beta$ for all diffusion maps-based methods, demonstrating that main approach using the multiscale graph correlation is the most powerful method against varying dependency structure.

\subsection{Degree-Corrected Stochastic Block Model}
In this section we compare different embeddings under the degree-corrected stochastic block model, which better reflects many real-world networks. The degree-corrected stochastic block model is an extension of stochastic block model by introducing an additional random variable $c_{i}$ to control the degree of each node. 

We set $n=200$ with two blocks, select the binary block-membership $Z_i$ uniformly in $\Omega=\{0,1\}$, and generate the edge probability by 
\begin{equation}
\mc{A}(i,j) \mid Z_{i}, Z_{j}, C_{i}, C_{j}  \sim Bernoulli \{ 0.2 C_{i} C_{j} \cdot \mathbb{I}( |Z_{i} - Z_{j}| = 0 ) + 0.05 C_{i} C_{j} \cdot \mathbb{I}(|Z_{i} - Z_{j}| = 1) \},
\label{eq:tau}
\end{equation} 
where $C_{i} \overset{i.i.d.}{\sim} Uniform(1 - \tau, 1 + \tau)$ for $i = 1, \ldots, n$, and $\tau \in [0, 1]$ is a parameter to control the amount of variability in the edge degree, e.g., as $\tau$ increases, the model becomes more complex as the variability of the edge probability becomes larger; when $\tau=0$, the above model reduces to a two-block stochastic block model without any variability induced by $\{ C_{i} : i=1,2,\ldots,n\}$. 
We again generate the nodal attributes $\mc{X}$ as a noisy version of the true block-membership via Bernoulli distribution, i.e., for each $i$, $X_{i}= Z_{i}$ with probability $0.6$, and equals the wrong label with probability $0.4$.  Figure~\ref{fig:powerplot}(b) compares different embedding choices using multiscale graph correlation.

%%%%%%%%%%%%%%%%%%%%%
\subsection{Random Dot Product Graph Simulations}
\label{ssec:RDPG}
Next we present a variety of random dot product graph simulations by generating the latent variables via the $20$ relationships in \citet{shen2017mgc} with different levels of noise, consisting of various linear, monotonic and non-monotonic relationships. The details of simulation schemes are in the Supplementary Material, and a general outline for data generating process is:
\begin{align}
\label{eq:RDPG_general}
\left( \begin{array}{cc} \tilde{W}_{i} & \tilde{X}_{i} \end{array} \right) & \overset{i.i.d.}{\sim} F_{\tilde{W} \tilde{X}} \qquad i = 1, 2, \ldots, n, \nonumber \\	
\mc{A}(i,j) \mid W_{i}, W_{j} & \sim Bernoulli\left( \langle W_{i}, W_{j} \rangle \right),~i<j=1,2,\ldots,n,
\end{align}
where $W_{i} = \{ \tilde{W}_{i} - \min( \{ \tilde{W}_{j} : j=1,2,\ldots,n \} ) \} / \{ \max(\{ \tilde{W}_{j} : j=1,2,\ldots,n \}) - \min(\{ \tilde{W}_{j} : j=1,2,\ldots,n \} \}$ for $i=1,2,\ldots, n$, so that all the latent variable range from 0 to 1. We apply the same scaling from $\tilde{X}_i$ to $X_i$ for visual consistency. 
\begin{figure}[H]
	\centering
	\includegraphics[width=0.8\textwidth]{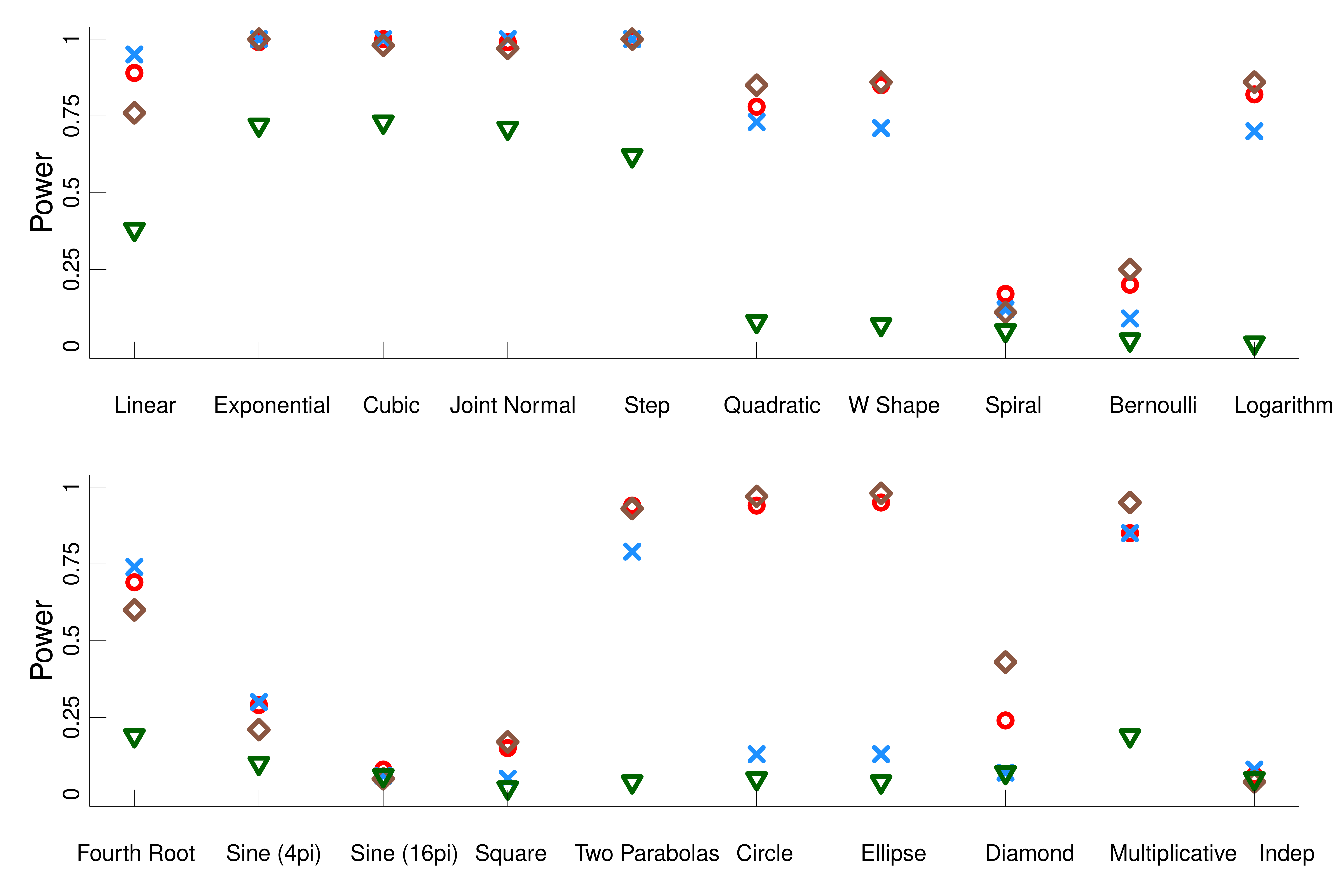}
	\caption{ \label{fig:RDPG} Power comparison for 20 different random dot product graphs with $n=50$ nodes per $m=100$ independent replicates. It shows that when latent positions $W_i$ and nodal attributes $X_i$ are dependent via a close-to-linear relationship at upper panel, diffusion multiscale graph correlation (red circle), diffusion distance correlation (blue X), and diffusion Heller-Heller-Gorfine (brown diamond) achieve similar power while the Fosdick-Hoff test (green triangle) is slightly worse due to its model-based nature. When non-linearity between $W_i$ and $X_i$ becomes evident like circle or ellipse at lower panel, multiscale graph correlation and Heller-Heller-Gorfine are the two best performing correlation measure, which is somewhat consist with the empirical results in \cite{shen2017mgc} for non-network data. }
\end{figure}
Thus the latent positions and nodal attributes are correlated via a joint distribution of $F_{\tilde{W} \tilde{X}}$, including linear, quadratic, circle, etc. 
Figure~\ref{fig:RDPG} shows empirical power obtained from $m=100$ independent replicates when the number of nodes is $n=50$, %A lack of power in the Fosdick-Hoff test is evident even though data generative model in Equation~\ref{eq:RDPG_general} agrees with \citet{fosdick2015testing}'s. 
for which all the diffusion map-based methods work fairly well. Note that the last scenario is an independent relationship and all tests achieve a power approximately at $0.05$, implying that they are all valid tests; there are also a few dependencies of very low power due to the complexity of the relationship, e.g., sine, spiral, square, etc., but their powers all converge to $1$ as sample size $n$ increases.
%%%%%%%%%%%%%%%%%%%%%%%%%%%%%%%%%%%%%%%%
\section{Graph Embedding using Diffusion Multiscale Graph Correlation}
\label{ssec:dis}

This section demonstrates that in deriving the diffusion correlation, we preserve dependency structure between $\mc{A}$ and $\mc{X}$ without cross-validation or over-fitting by virtue of effectively estimating parameters of $t$ and $q$. As a reminder, the dimension choice $q$ is selected by the second elbow of the absolute eigenvalue scree plot via the profile likelihood method from~\citet{ZhuGhodsi2006}. The choice of $t^{*}$ is based on a smoothed maximum. Viewed in another way, diffusion multiscale graph correlation selects the optimal diffusion map that maximizes the multiscale graph correlation. Thus any testing advantage shall come down to whether it is able to optimize the embedding without over-fitting, and we investigate how well our procedure is able to preserve the dependency compared to adjacency spectral embedding.
\begin{figure}[ht]
	\centering
	\begin{minipage}[b]{0.20\textwidth}
		\includegraphics[width=\textwidth]{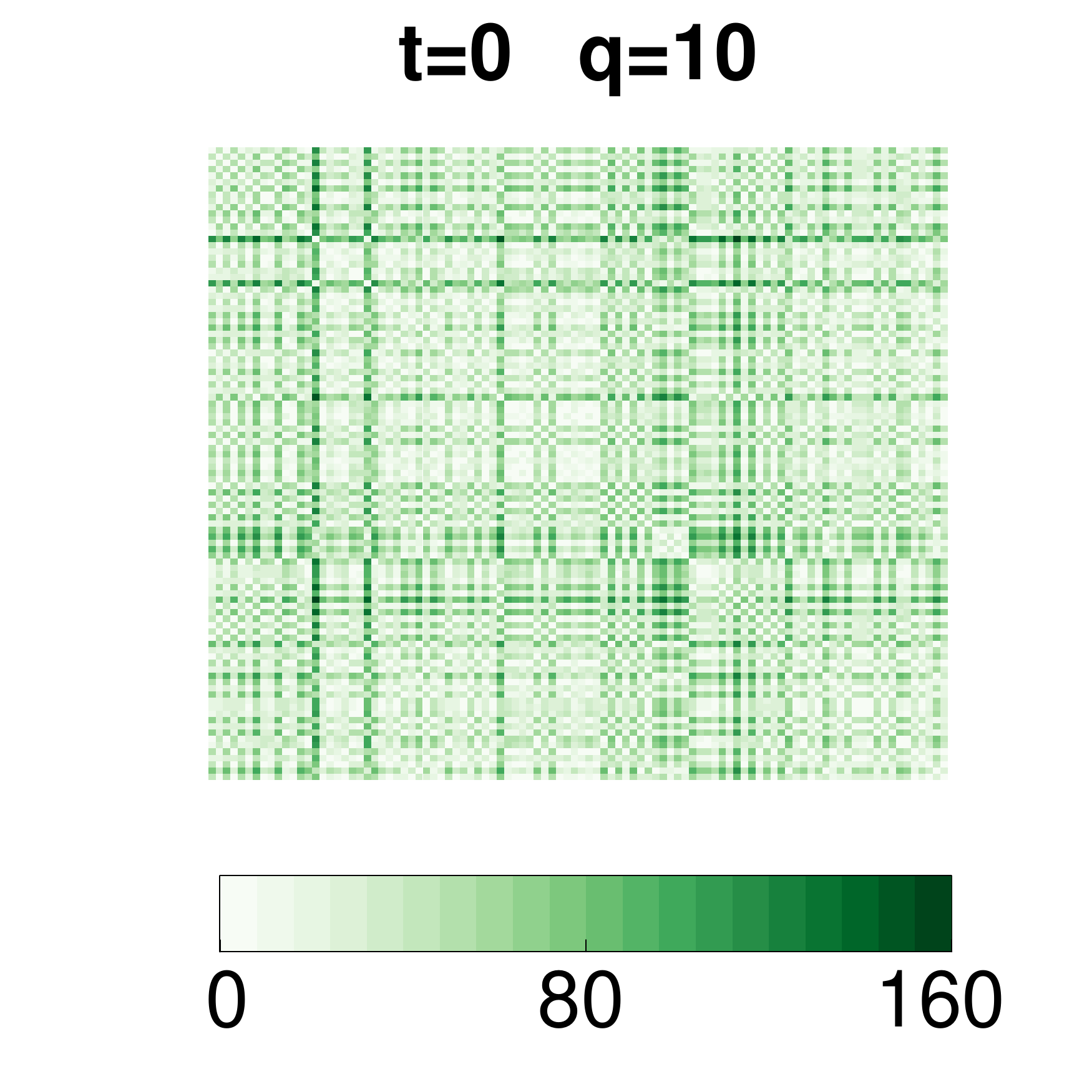} \\
		\centering (a)
	\end{minipage}
	\begin{minipage}[b]{0.20\textwidth}
		\includegraphics[width=\textwidth]{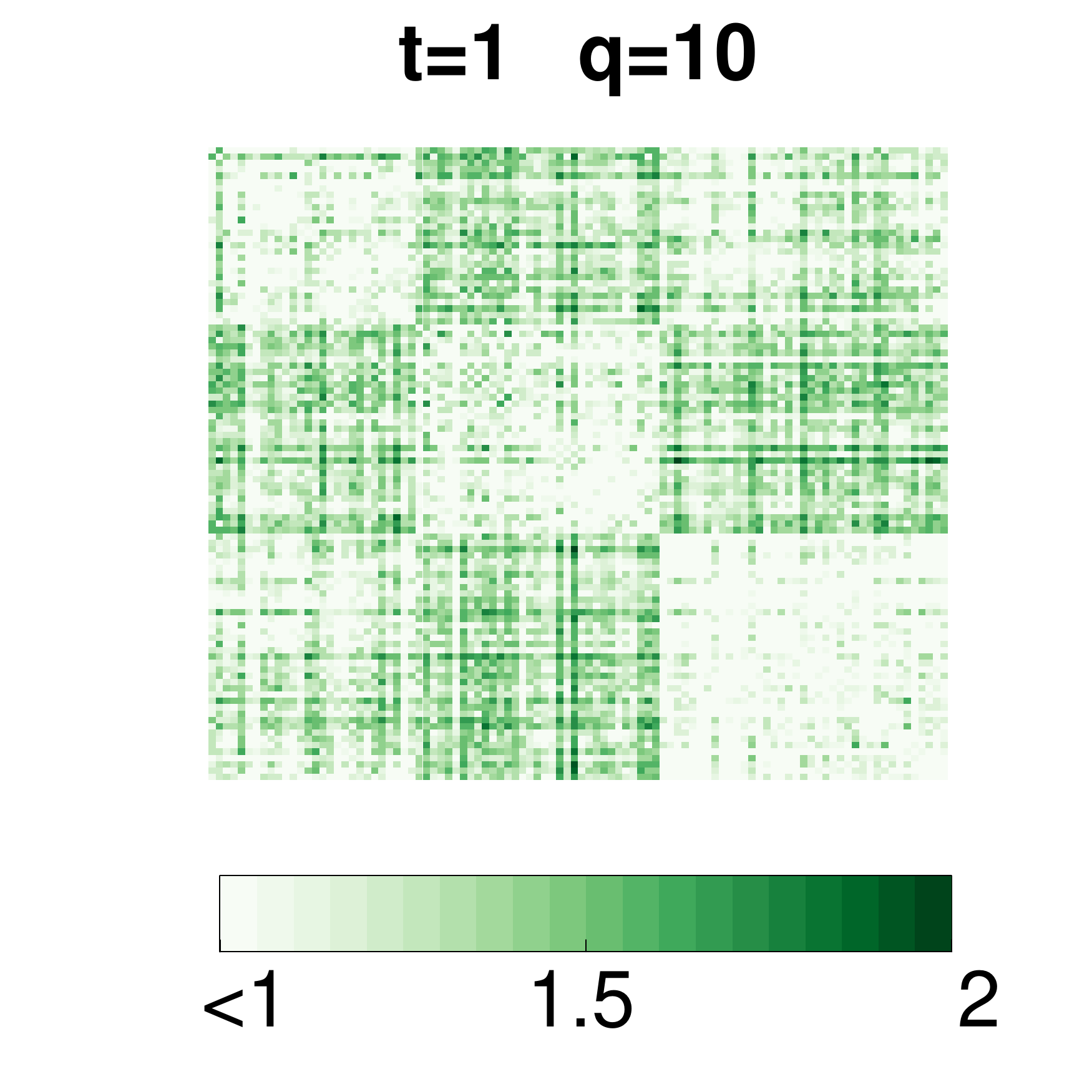} \\
		\centering (b)
	\end{minipage}
	\begin{minipage}[b]{0.20\textwidth}
		\includegraphics[width=\textwidth]{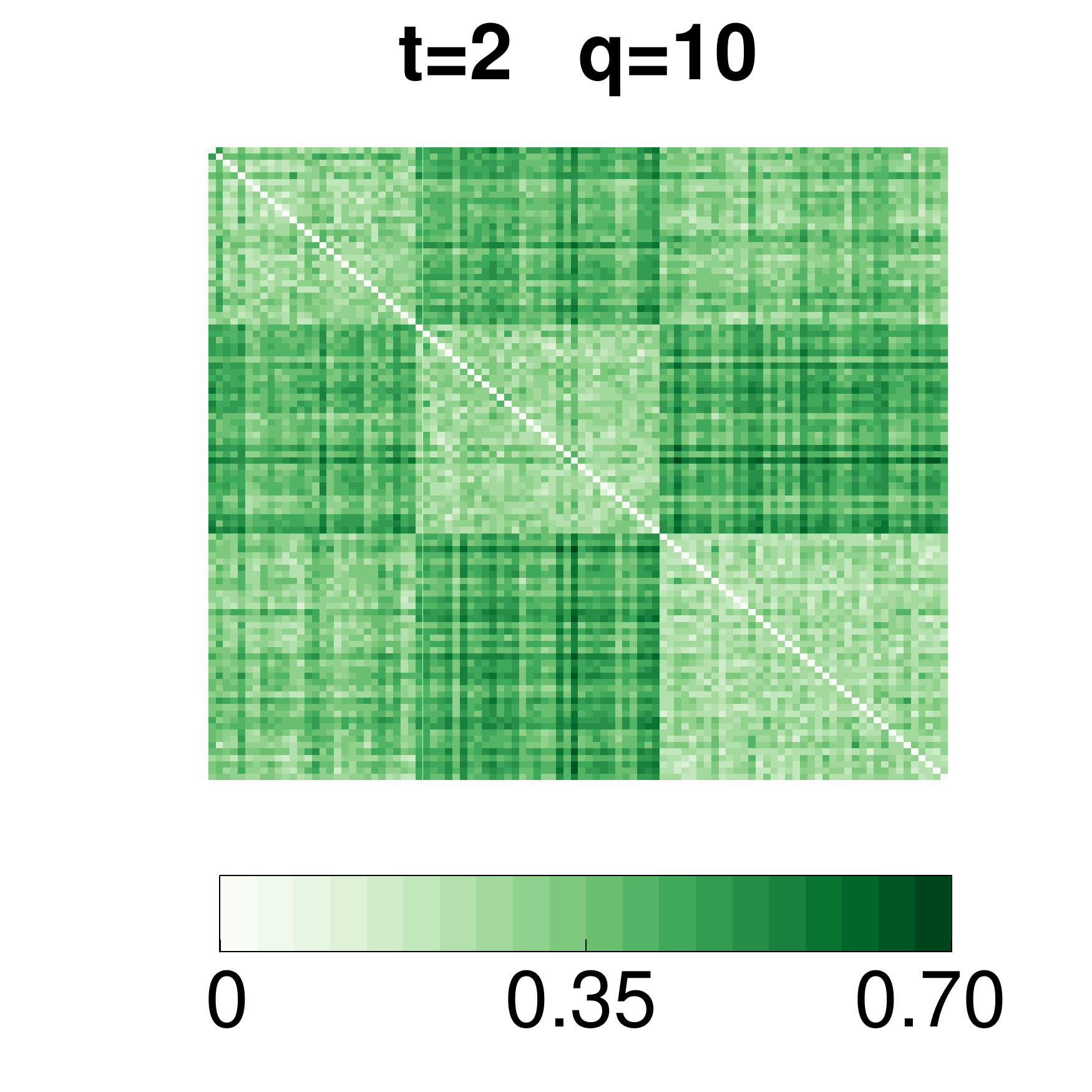} \\
		\centering (c)
	\end{minipage}
	\begin{minipage}[b]{0.20\textwidth}
		\includegraphics[width=\textwidth]{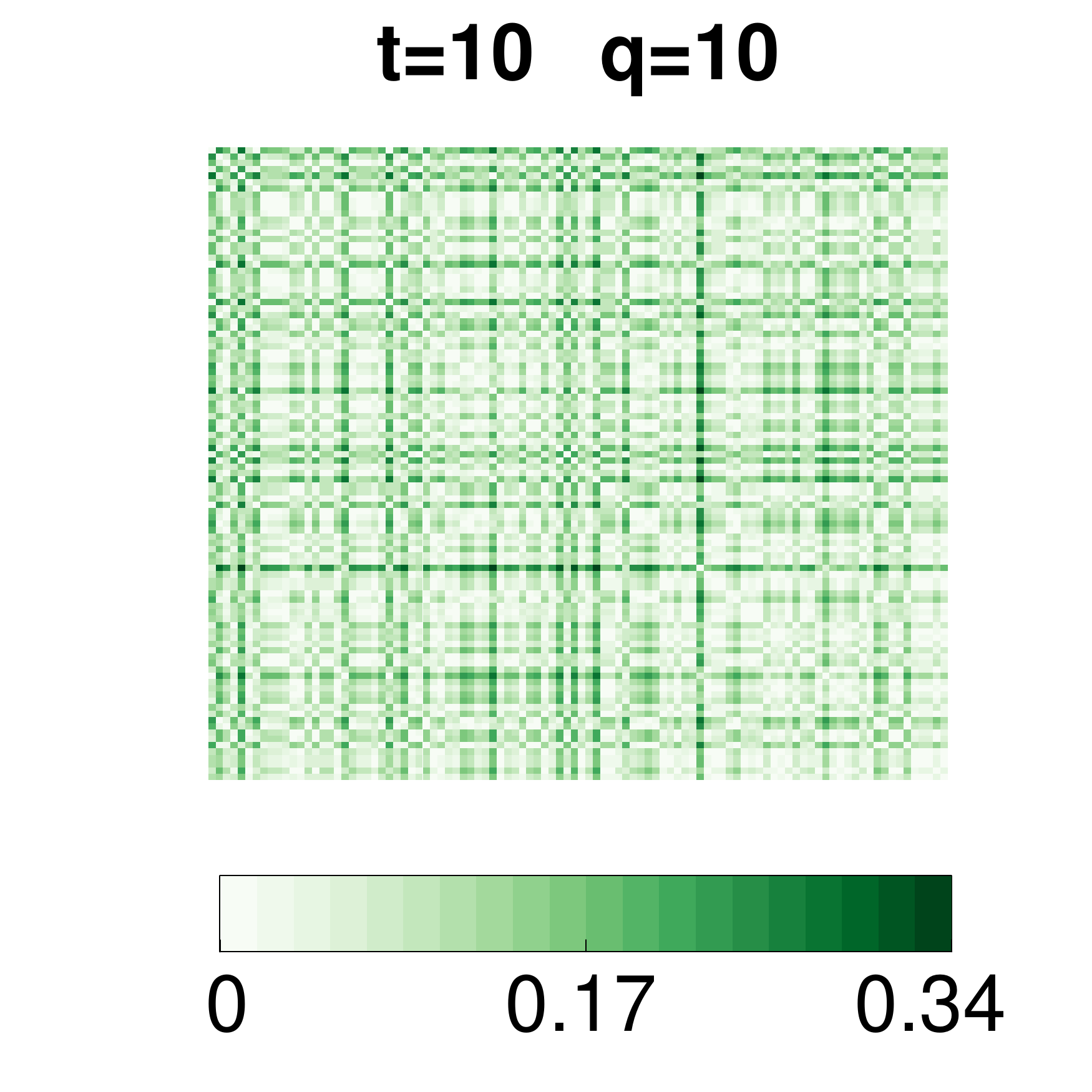} \\
		\centering (d)
	\end{minipage}
	\\ \bigskip
	\begin{minipage}[b]{0.20\textwidth}
		\includegraphics[width=\textwidth]{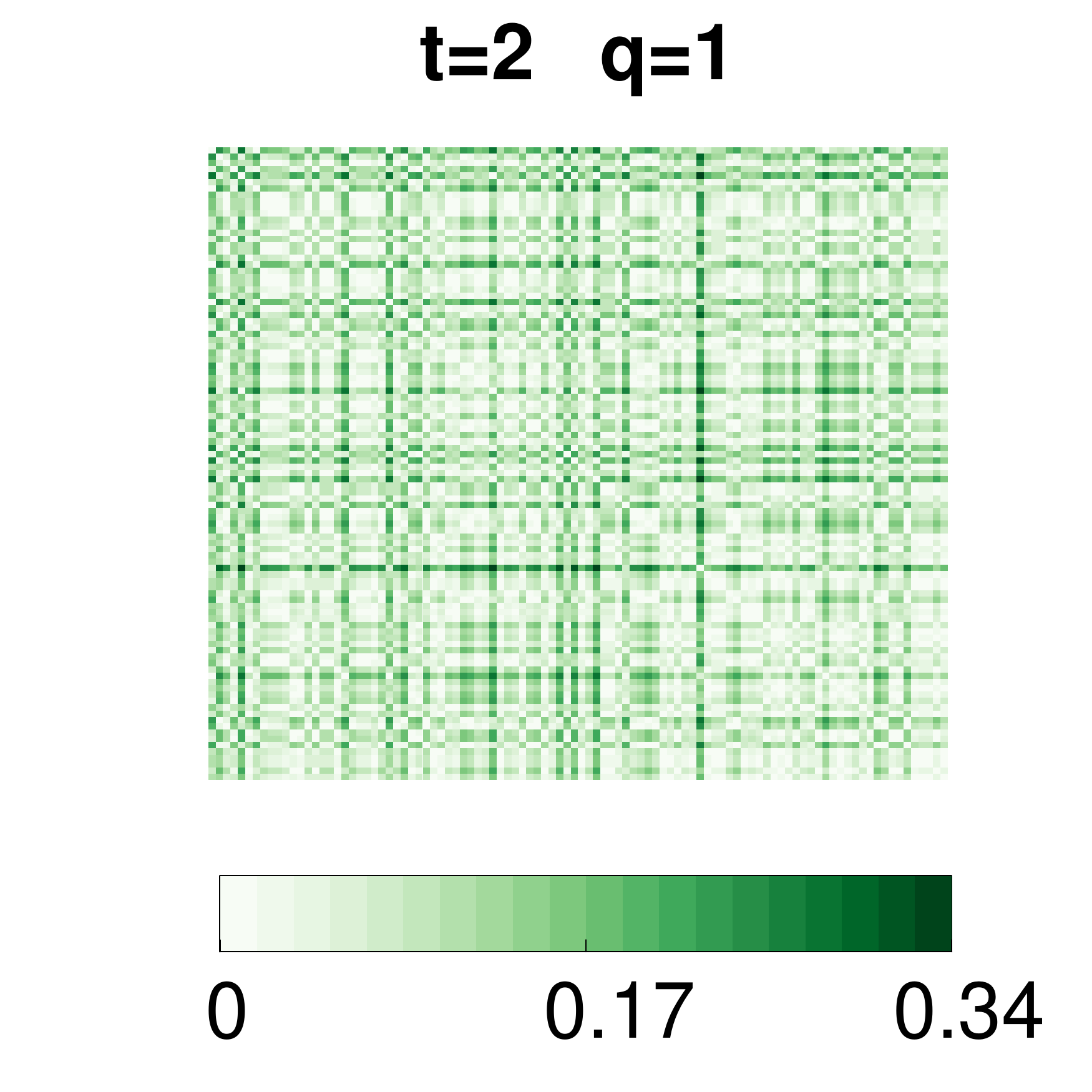} \\
		\centering (e)
	\end{minipage}
	\begin{minipage}[b]{0.20\textwidth}
		\includegraphics[width=\textwidth]{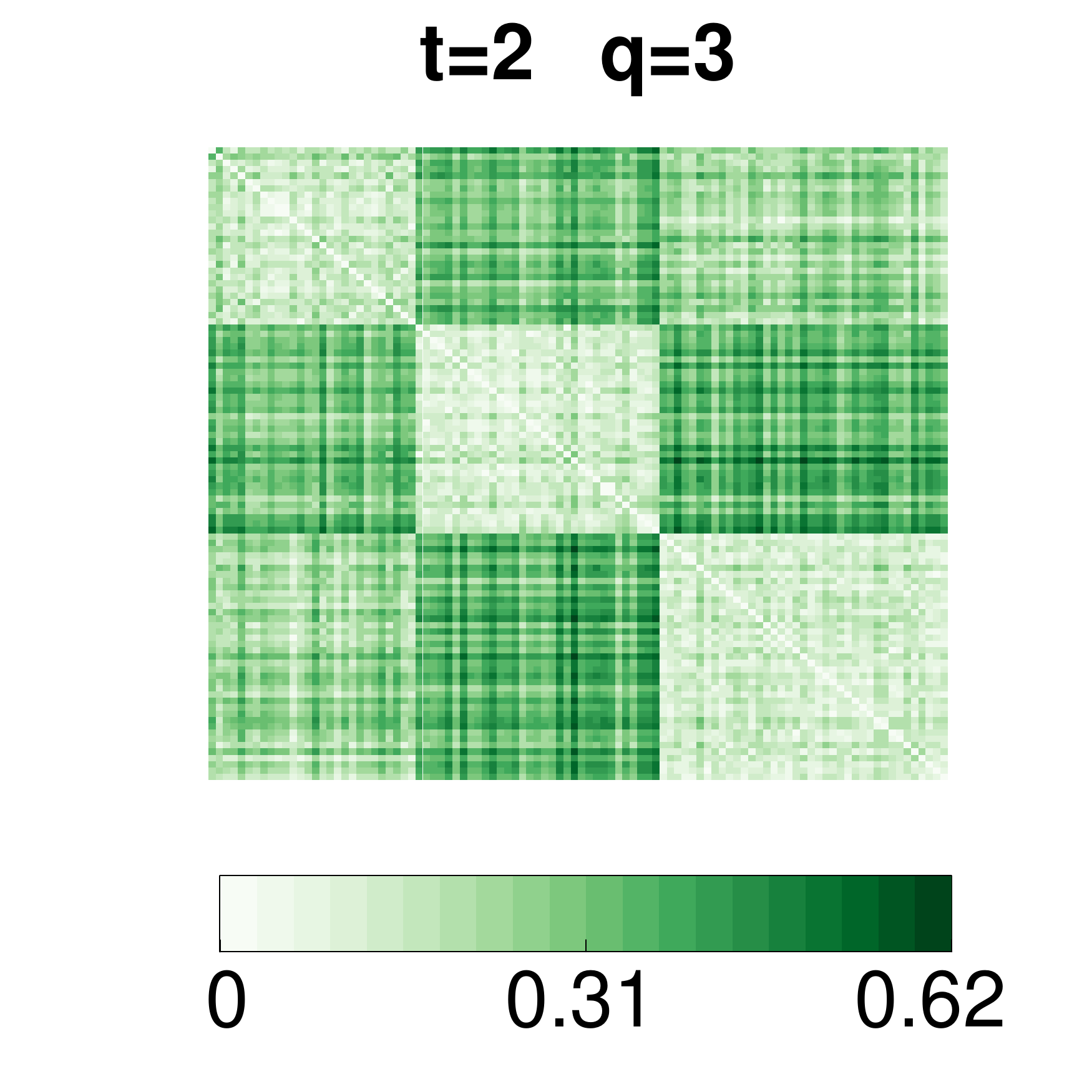} \\
		\centering (f)
	\end{minipage}
	\begin{minipage}[b]{0.20\textwidth}
		\includegraphics[width=\textwidth]{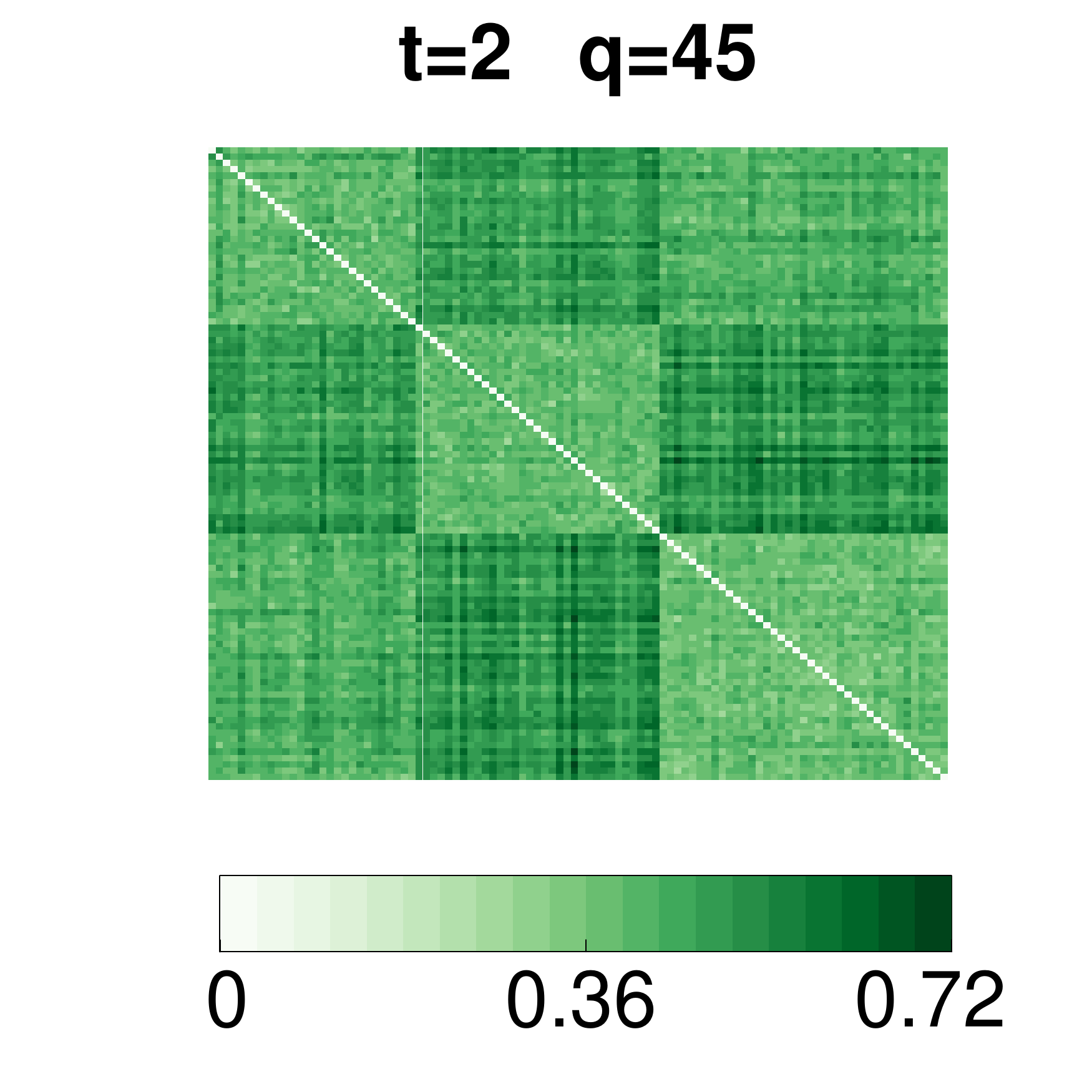} \\
		\centering (g)
	\end{minipage}
	\begin{minipage}[b]{0.20\textwidth}
		\includegraphics[width=\textwidth]{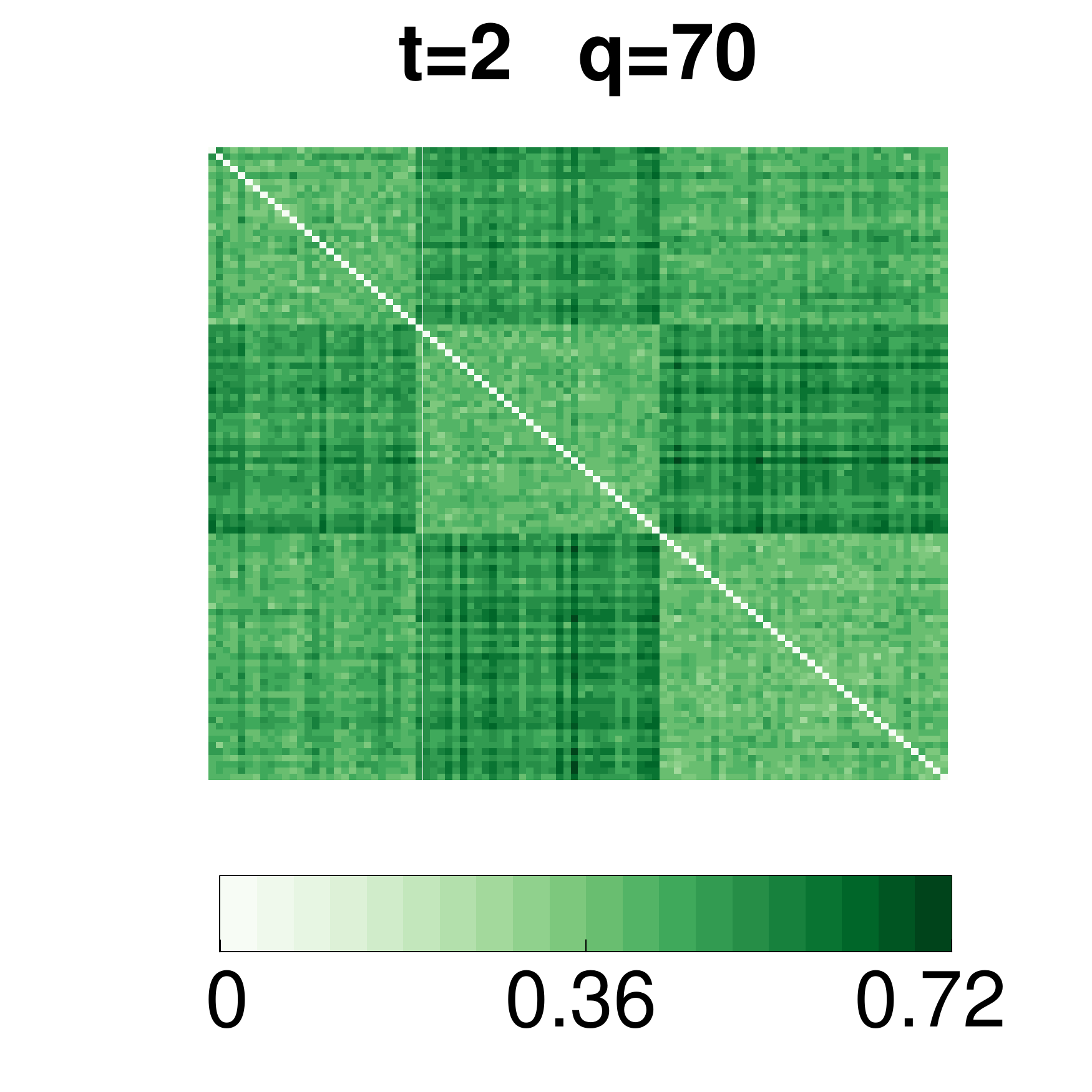} \\
		\centering (h)
	\end{minipage}
	\caption{Generate a three-block adjacency matrix $\mc{A}$ by Equation~\ref{eq:Three} at $n=100$, and compute the diffusion distances at each combination of $(t,q)$. A visualization of adjacency matrix is provided in Fig.~\ref{fig:embedding} (a); upon fixing a good $t$, many choices of $q$  preserve the block structure. Note that the first three elbows of eigenvalues are $(1,45,70)$ and $t^{*}=2$, so panel (g) is the optimal diffusion map by diffusion multiscale graph correlation.}
	\label{fig:diffusions}
\end{figure} 

\begin{figure}[H]
	\centering
	\begin{minipage}[b]{0.20\textwidth}
		\includegraphics[width=\textwidth]{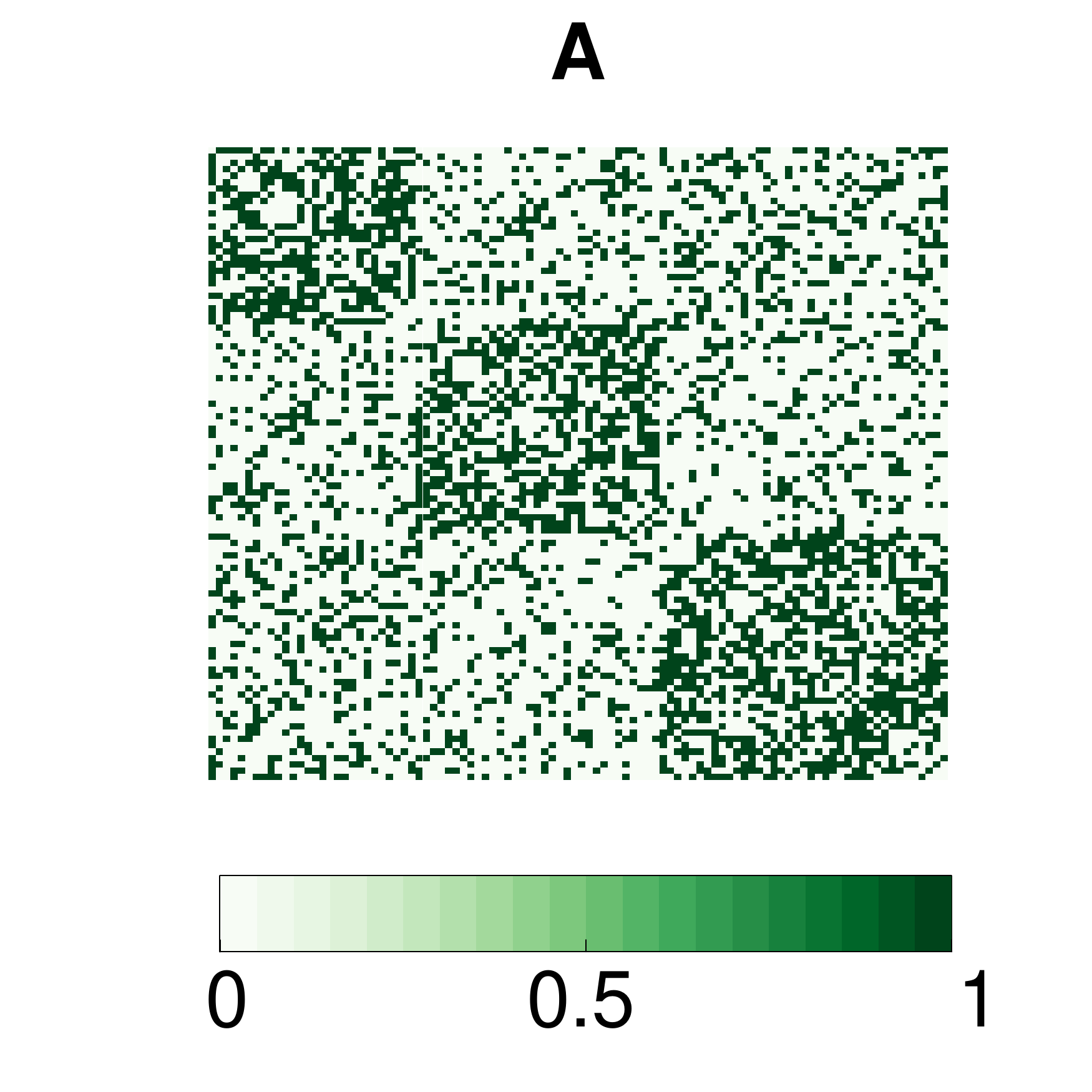} \\
		\centering (a)
	\end{minipage}
	\begin{minipage}[b]{0.20\textwidth}
		\includegraphics[width=\textwidth]{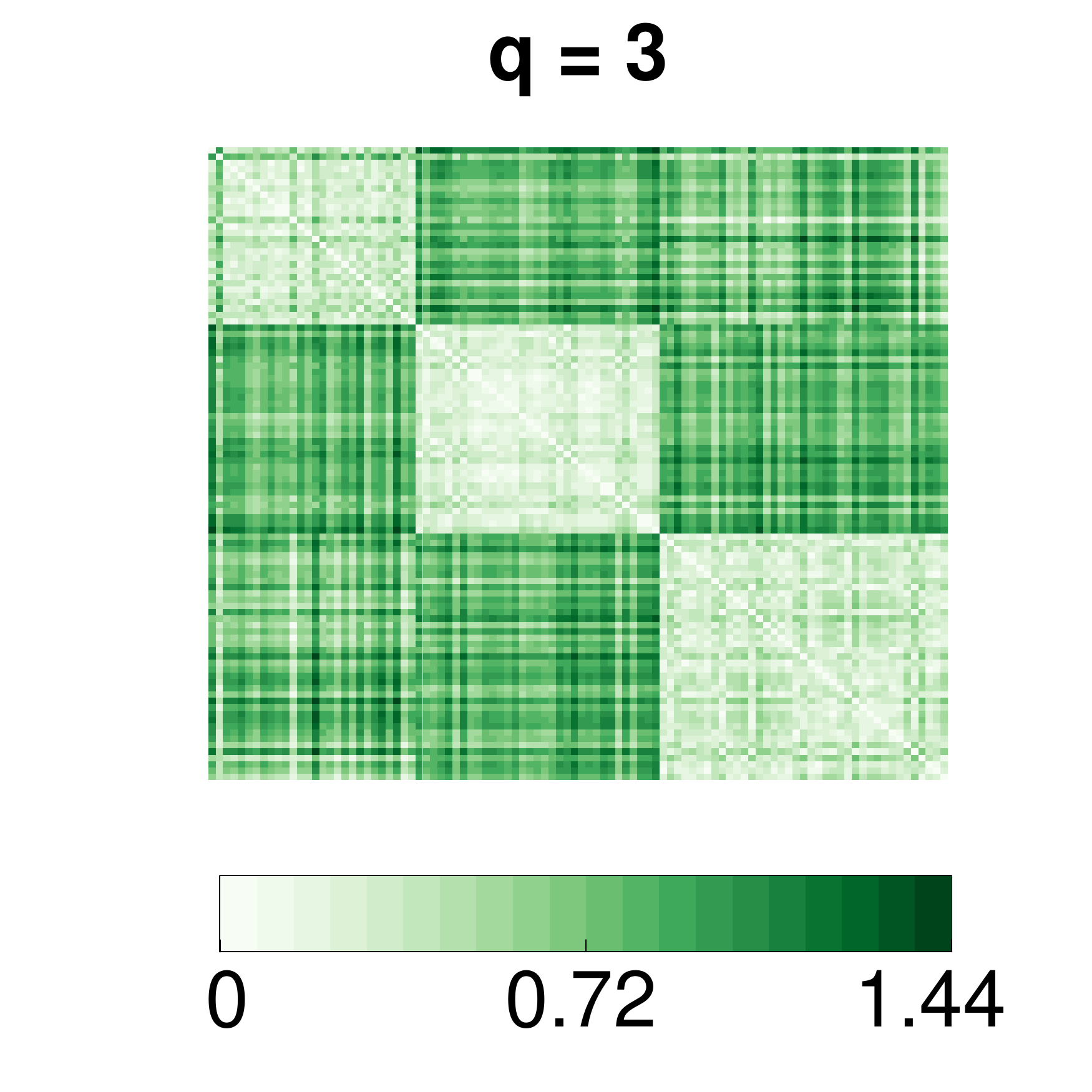} \\
		\centering (b)
	\end{minipage}
	\begin{minipage}[b]{0.20\textwidth}
		\includegraphics[width=\textwidth]{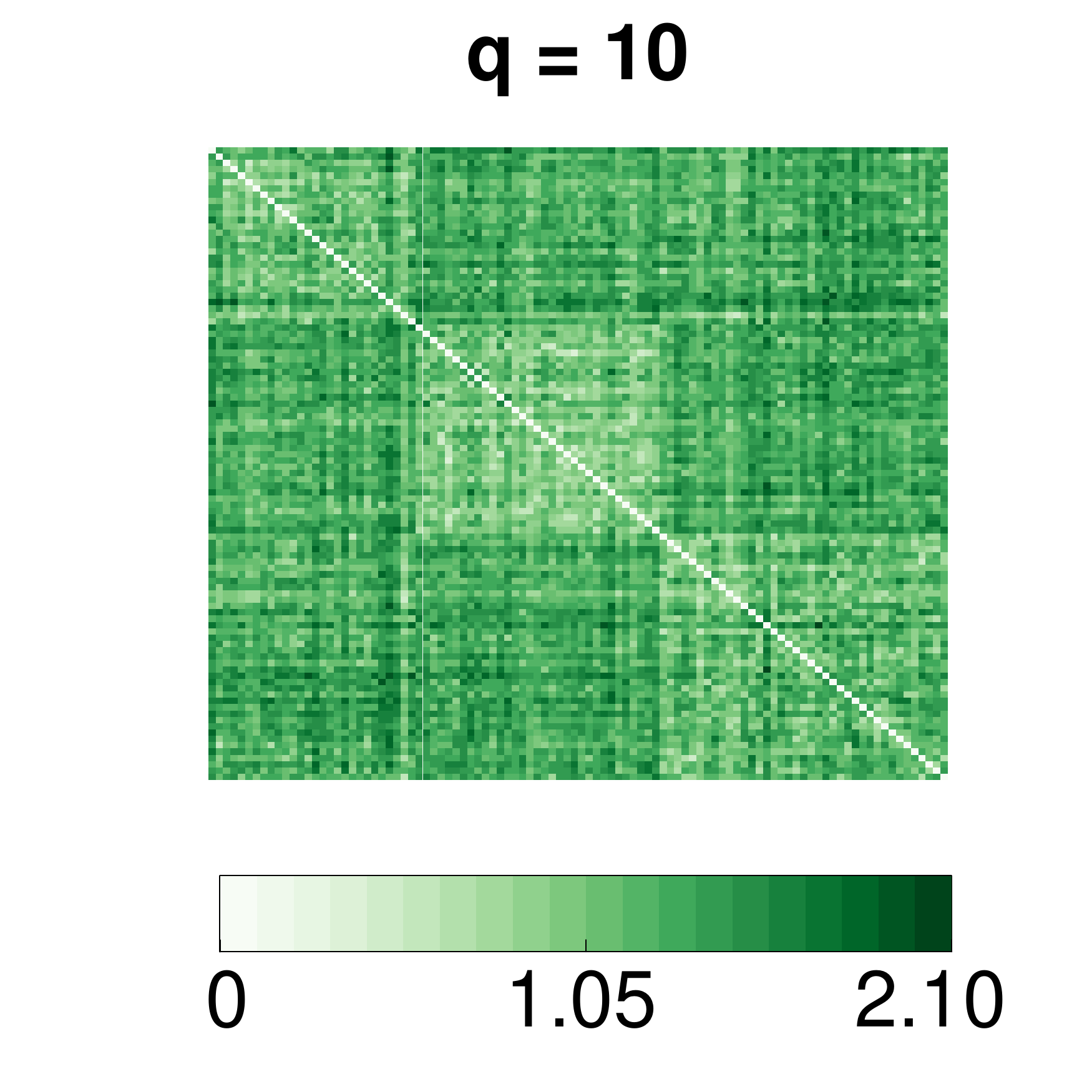} \\
		\centering (c)
	\end{minipage}
	\begin{minipage}[b]{0.20\textwidth}
		\includegraphics[width=\textwidth]{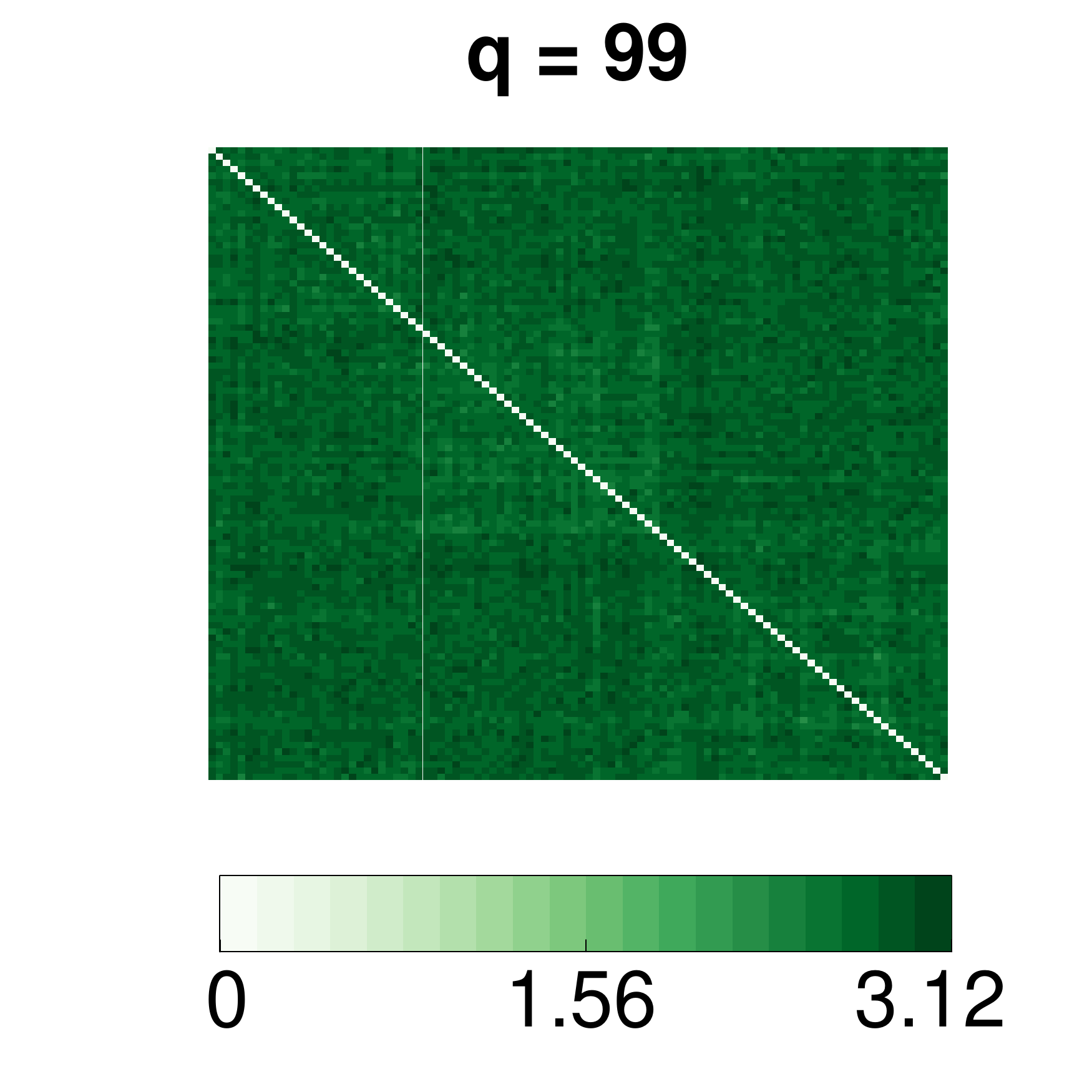} \\
		\centering (d)
	\end{minipage}
	\caption{Panel (a) shows the adjacency matrix of three-block adjacency matrix $\mc{A}$ generated by Equation~\ref{eq:Three}. Panel (b)-(d) show the Euclidean distance matrix of adjacency spectral embedding at increasing $q$, using the same adjacency matrix of Panel (a). Only adjacency spectral embedding at $q=3$, namely at the correct dimension, is able to display a clear block structure. Note that the first three elbows are $(1,45,70)$, so adjacency spectral embedding has a more obscure block structure when the dimension is chosen via the scree plot, comparing to the diffusion correlation-based embedding in Fig.~\ref{fig:diffusions} (g).}
	\label{fig:embedding}
\end{figure}
Figure~\ref{fig:diffusions} presents the diffusion distances at different $t$ and $q$ for the three-block stochastic block model in Equation~\ref{eq:Three}. 
Although the resulting embedding is sensitive to both $t$ and $q$ in Fig.~\ref{fig:diffusions} (a)--(d), at optimal $t^{*}=2$ it is robust against $q$, e.g., Fig.~\ref{fig:diffusions} (e)--(h) show that for a wide range of $q$ the block structure is preserved in the resulting diffusion maps including the second elbow, so the diffusion correlation-based embedding preserves the dependency structure well.

On the other hand, Fig.~\ref{fig:embedding} shows the Euclidean distance of the adjacency spectral embedding \citep{SussmanEtAl2012} applied to the same adjacency matrix. For adjacency spectral embedding, the correct dimensional choice equals the number of blocks, i.e., the distance matrix at $q=3$ shows a clear block structure in Fig.~\ref{fig:embedding} (b). However, a slight misspecification of $q$ can cause the embedding to have a more obscure block structure, and the elbow method often fails to find the correct $q$ for adjacency spectral embedding.  

\begin{figure}[H]
	\centering
	\begin{minipage}[b]{0.49\textwidth}
		\includegraphics[width=\textwidth]{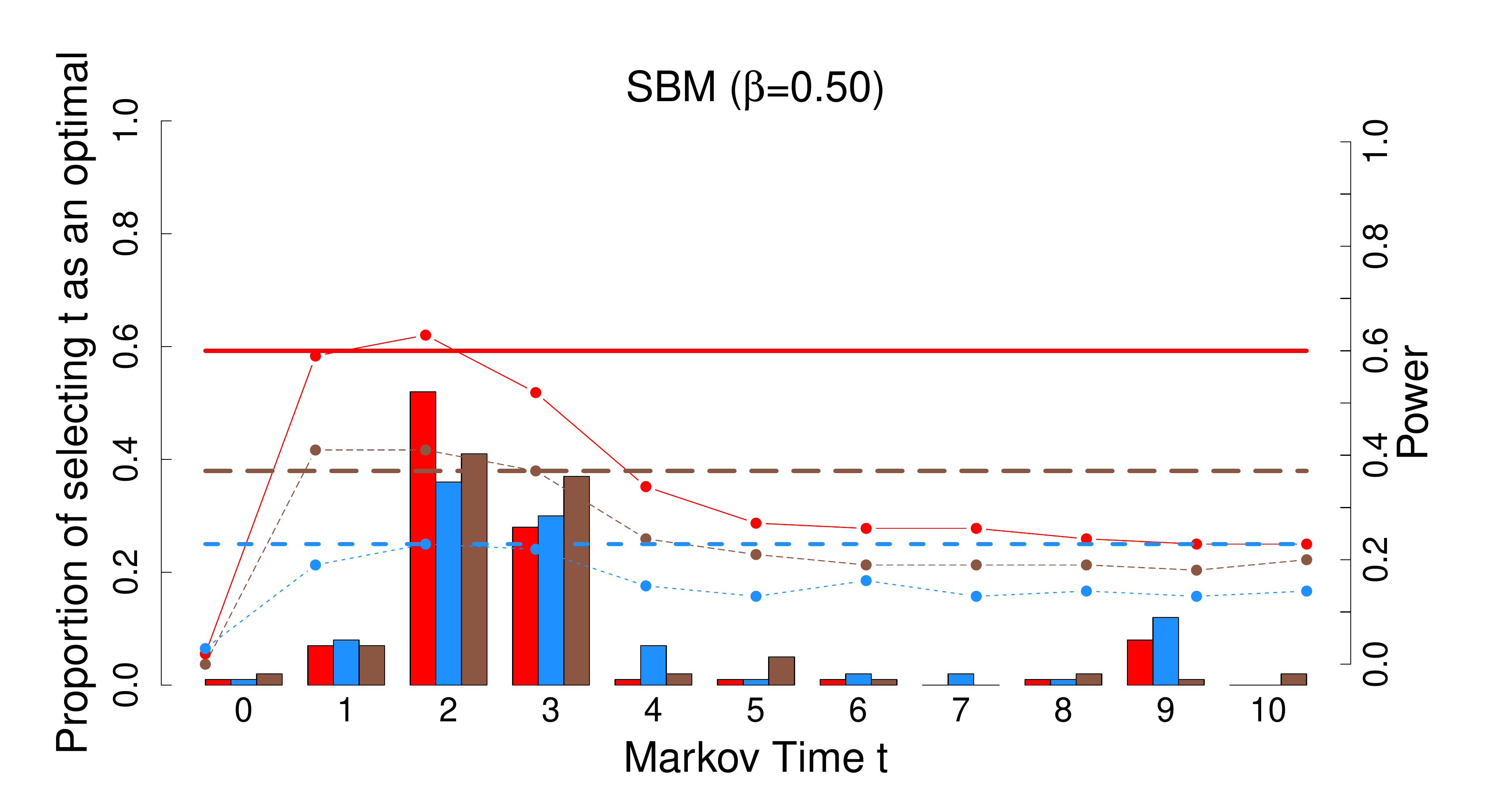}\\
		\centering (a)
	\end{minipage}
	\begin{minipage}[b]{0.49\textwidth}
		\includegraphics[width=\textwidth]{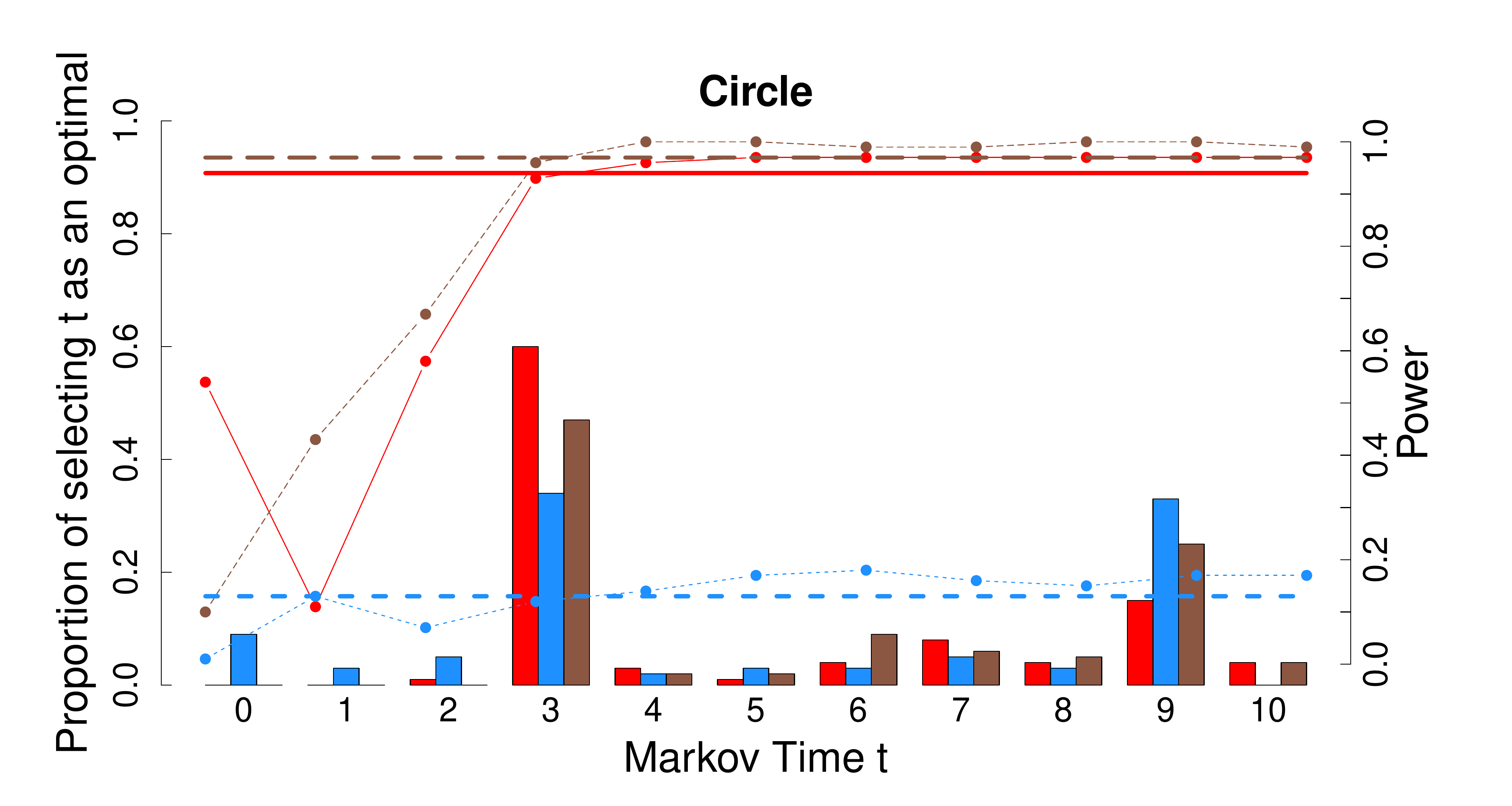} \\
		\centering (b)
	\end{minipage}
	\caption{ 
		\label{fig:timeplot} 
		Testing power comparison between diffusion \MGC, and~\MGC~on each diffusion map. Using $m = 100$ replicates, the solid red line plots the power of $\MGC^{*}_{n}(\{\mc{U}^{t}\},\mc{X})$; the dash line plots the power of $\MGC_{n}(\mc{U}^{t},\mc{X})$ for $t \in \{ 0,1,2,\ldots,10 \}$; the bar plot shows the proportion that diffusion \MGC~(solid) selects each $t \in \{ 0,1,2,\ldots,10 \}$ as the optimal $t^{*}$. Diffusion~\HHG~(small dashes) and diffusion~\dCorr~(dashes) are also added by different colors. For each method, the diffusion statistic is able to achieve an excellent power that is almost equivalent to the best possible power among all $t$. It suggests that the smoothed maximum scheme is able to identify the graph embedding that best preserves the dependency structure.} 
\end{figure}

Next we compare testing performance of the diffusion correlation-based embedding $\mc{U}^{t^{*}}$ versus all other diffusion maps $\mc{U}^{t}$. 
Figure~\ref{fig:timeplot} shows the proportion of choosing $t$ as the optimal among $\{0,1,2, \ldots, 10\}$ and the testing power for each $t$ and also $t^{*}$. Figure~\ref{fig:timeplot} (a) illustrates that under the stochastic block model dependency structure in Equation~\ref{eq:mono} with $\beta = 0.50$, diffusion multiscale graph correlation is mostly likely to choose $t^{*} = 2$ as the optimal time-step, and the testing power is almost equivalent to the best power among all $t \in \{0,1,2, \ldots, 10 \}$. The same phenomena hold for other diffusion correlations, and Fig.~\ref{fig:timeplot} (b) illustrates the results via the random dot product graph simulation example by Equation~\ref{eq:RDPG_general}. 

\section{Real Data Application}
\label{sec:realdata}
\begin{figure}[H]
	\centering
	\includegraphics[width=0.6\textwidth]{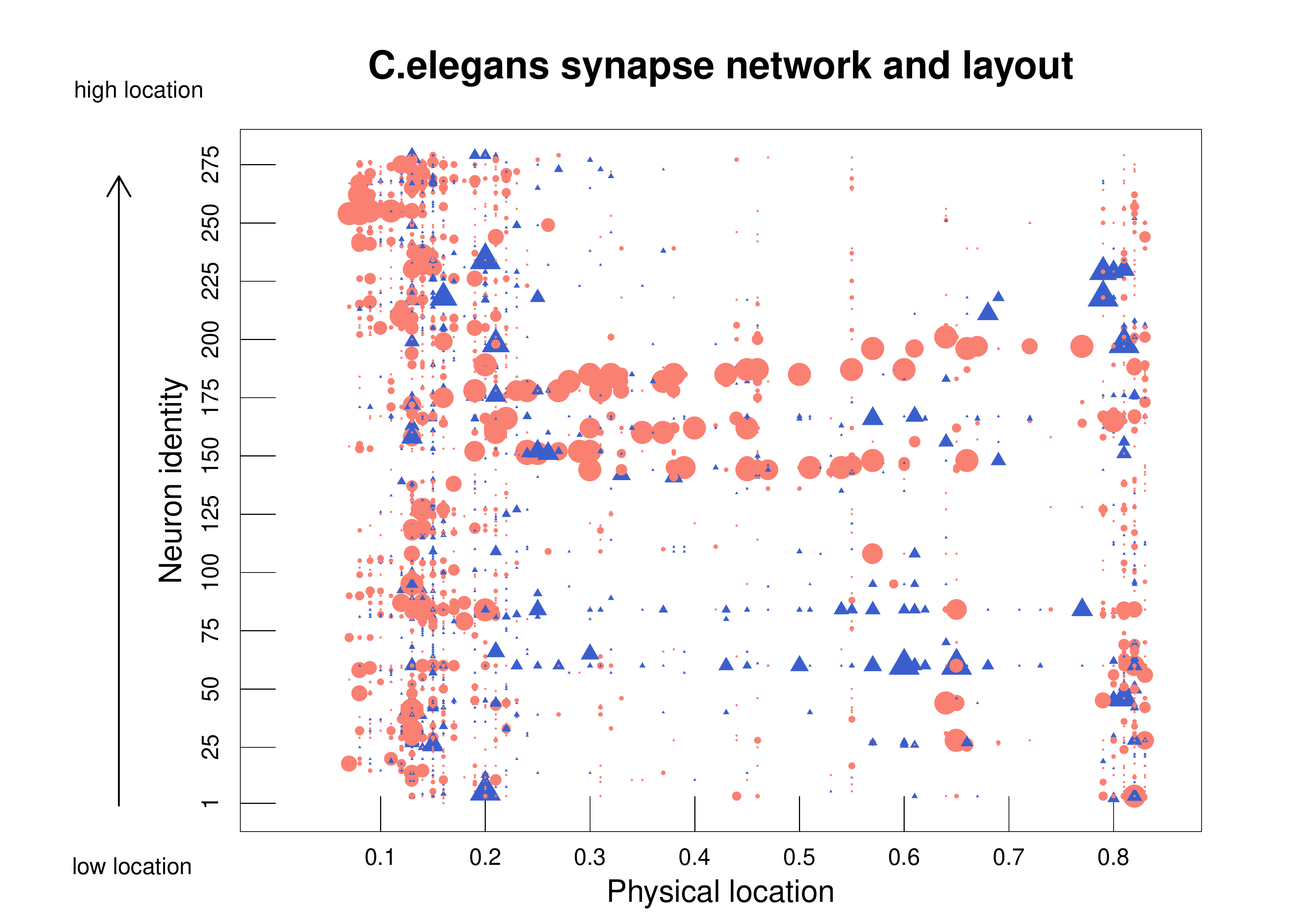} 
	\caption{\label{fig:celegans} Each dot at $(x,y)$ represents the existence of synapses from neuron at $y$ to neuron at $x$. Y-axis represents each neuron’s index assigned from low location to high location, and x-axis represents 68 different locations where neurons are positioned. Color of dots represents synapse type, either chemical (red circle) or electrical (blue triangle), and size of dots is proportional to the number of synapse but capped at 10.}
\end{figure}

As a real data example, we apply the methodology to the neuronal network of hermaphrodite \textit{Caenorhabditis elegans}, which composes of 279 nonpharyngeal neurons connecting each other through chemical and electrical synapses~\citep{varshney2011structural}. Each node represents an individual neuron, and each edge weight indicates the number of synapses between them. Among a few known attributes including types of neurotransmitter and role of neurons, we use one dimensional, continuous position of each neuron as the nodal attribute $\mc{X}$. Figure~\ref{fig:celegans} shows that neurons at low location and high location are connected to other neurons distributed throughout the region; while those at the relatively middle of location are connected to the neurons only within the narrower area.
The independence test between synapse connectivity and each neuron's position can be connected to growing number of studies on relationship between physical arrangement and functional connectivity in Caenorhabditis elegans~\citep{chen2006wiring, kaiser2006nonoptimal} and others'~\citep{cherniak2004global, alexander2012anatomical}. We binarize and symmetrize both chemical and electrical synapses, add them together to represent overall synapse connectivity of Caenorhabditis elegans, then apply diffusion multiscale graph correlation, diffusion distance correlation, diffusion Heller-Heller-Gorfine, and Fosdick-Hoff to test independence between connectivity through synapses and neuron's position. All methods result in similar significant p-values less than 0.002. 
\begin{figure}[ht]
	\centering
	\begin{minipage}[b]{0.20\textwidth}
		\includegraphics[width=\textwidth]{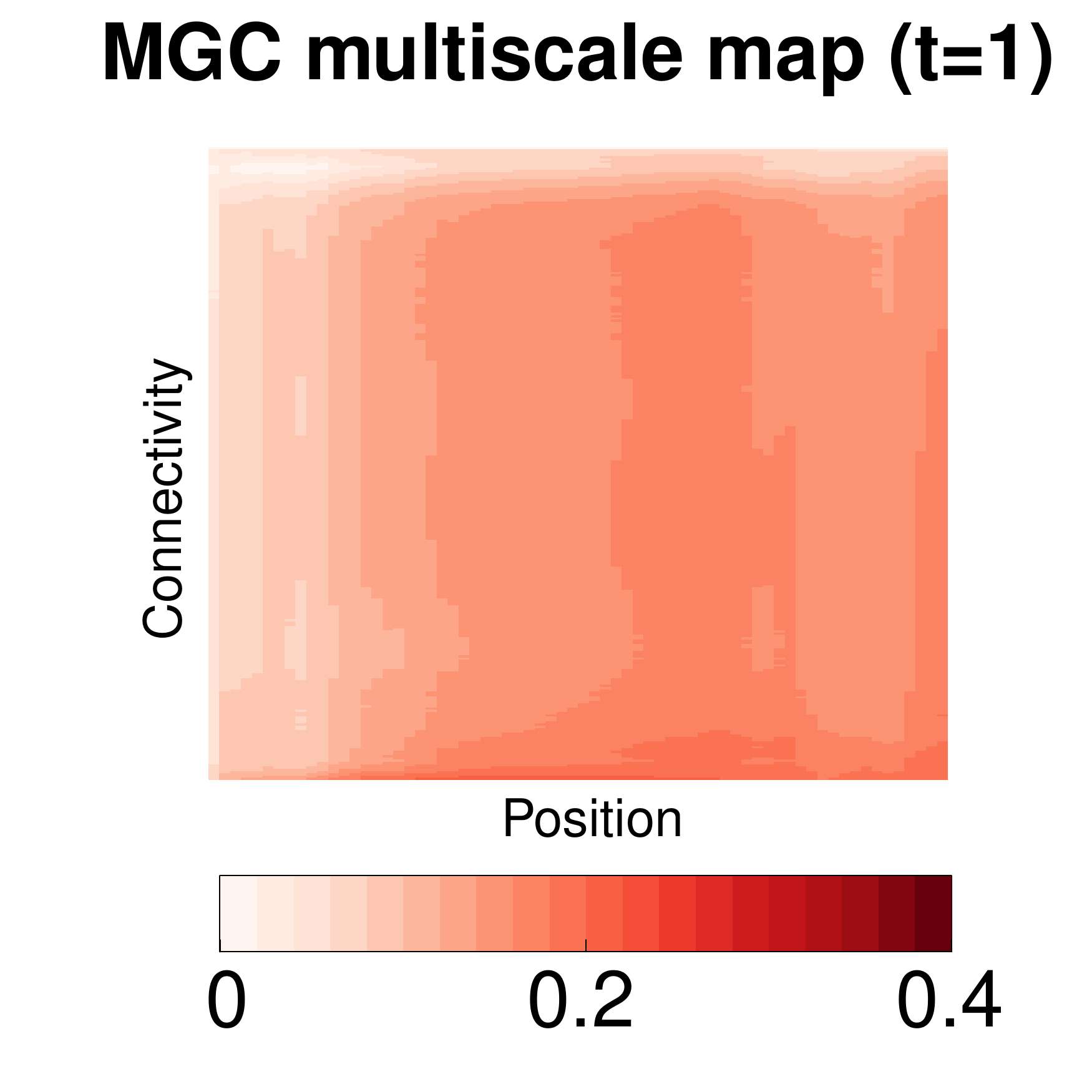} \\
		\centering (a)
	\end{minipage}
	\begin{minipage}[b]{0.20\textwidth}
		\includegraphics[width=\textwidth]{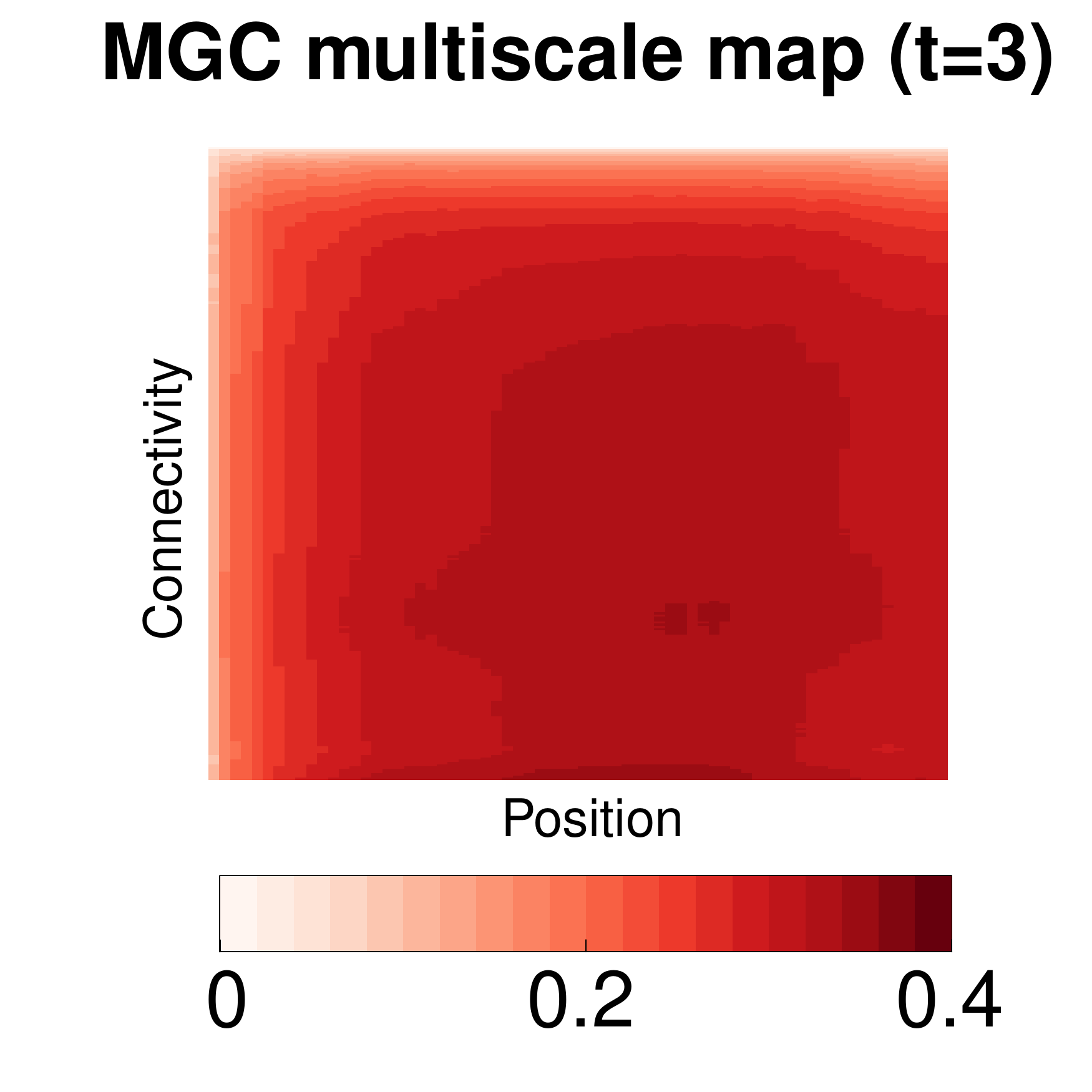} \\
		\centering (b)
	\end{minipage}
	\begin{minipage}[b]{0.20\textwidth}
		\includegraphics[width=\textwidth]{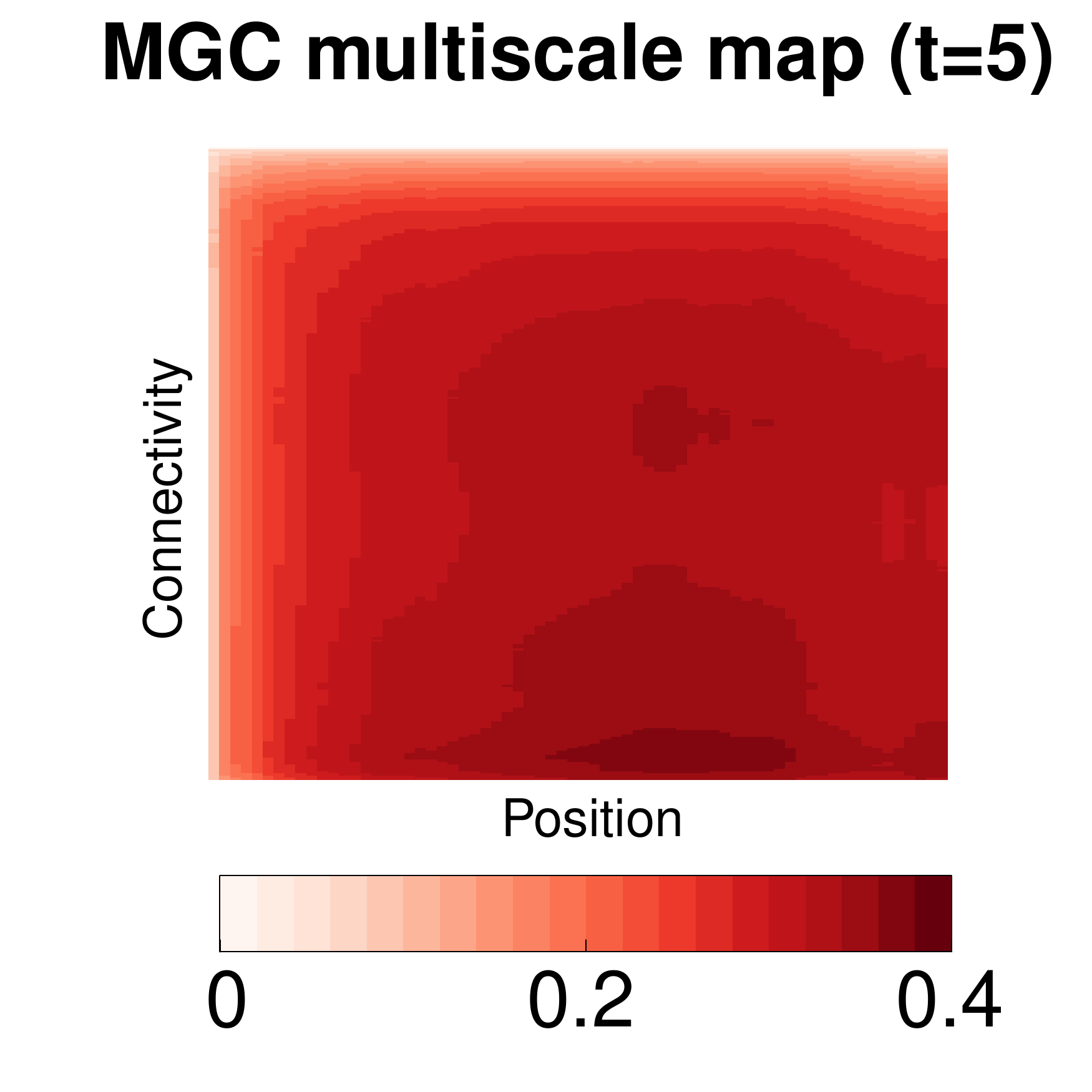} \\
		\centering (c)
	\end{minipage}
	\begin{minipage}[b]{0.20\textwidth}
		\includegraphics[width=\textwidth]{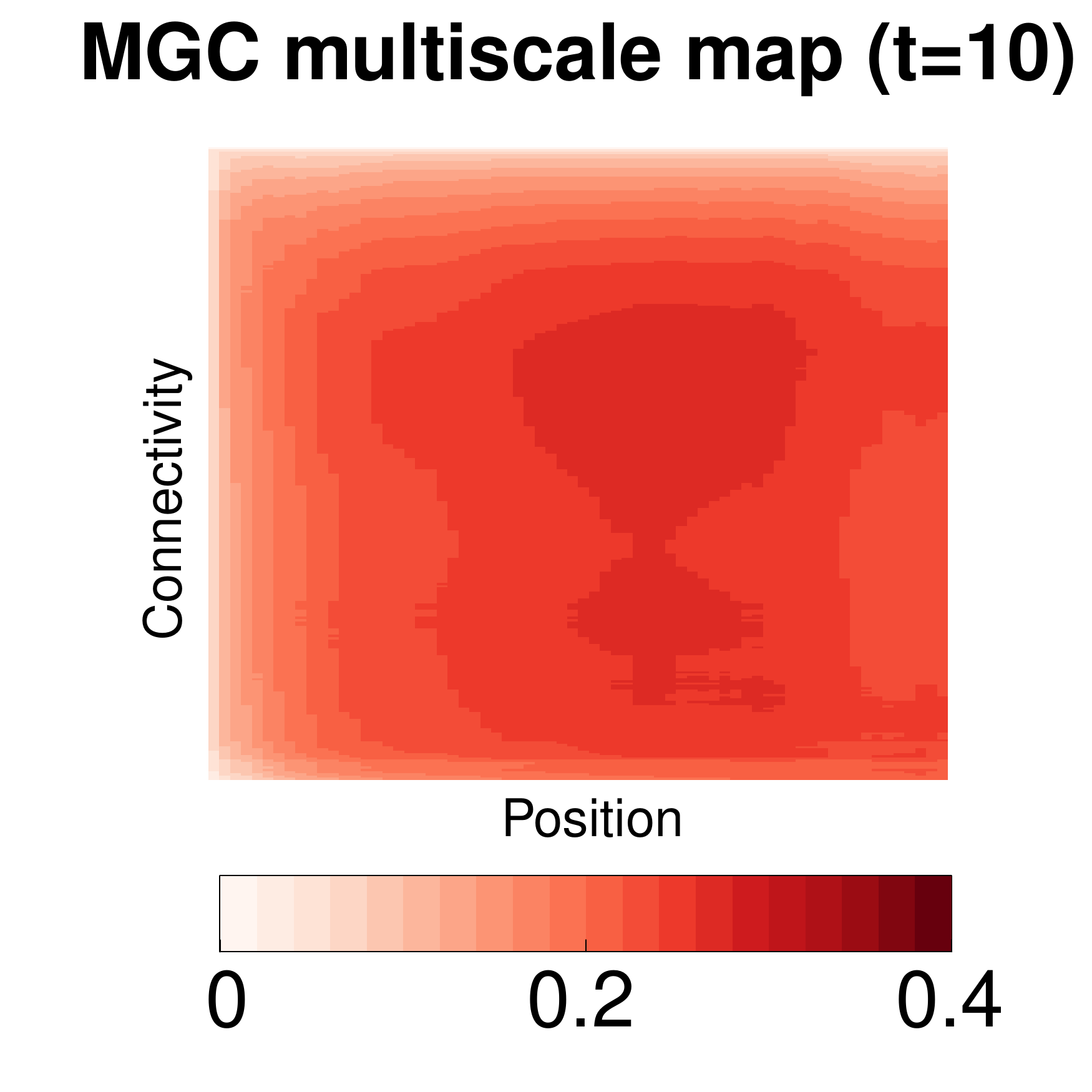} \\
		\centering (d)
	\end{minipage}
	\centering
	\begin{minipage}[b]{0.20\textwidth}
		\includegraphics[width=\textwidth]{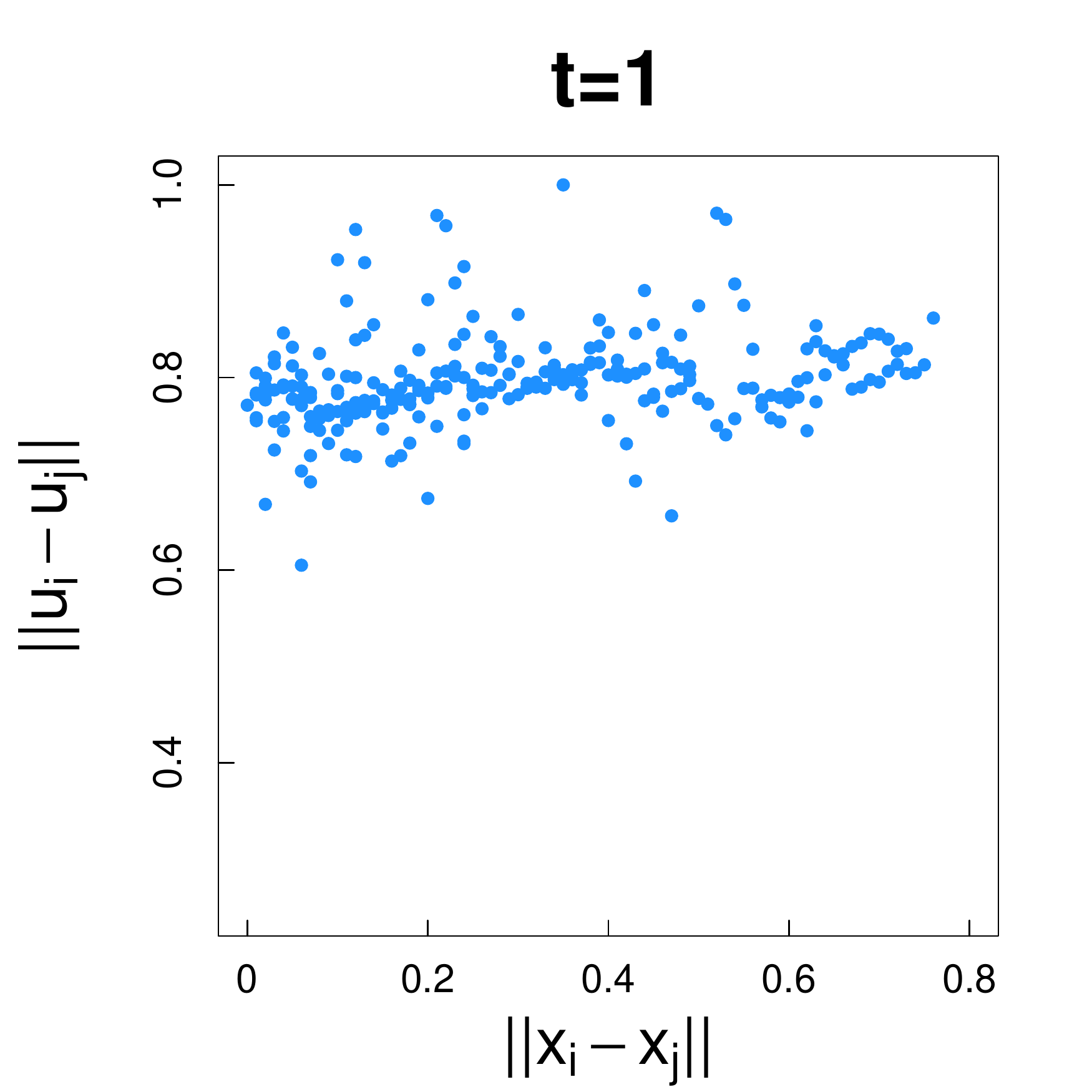} \\
		\centering (e) $\rho = 0.25$
	\end{minipage}
	\begin{minipage}[b]{0.20\textwidth}
		\includegraphics[width=\textwidth]{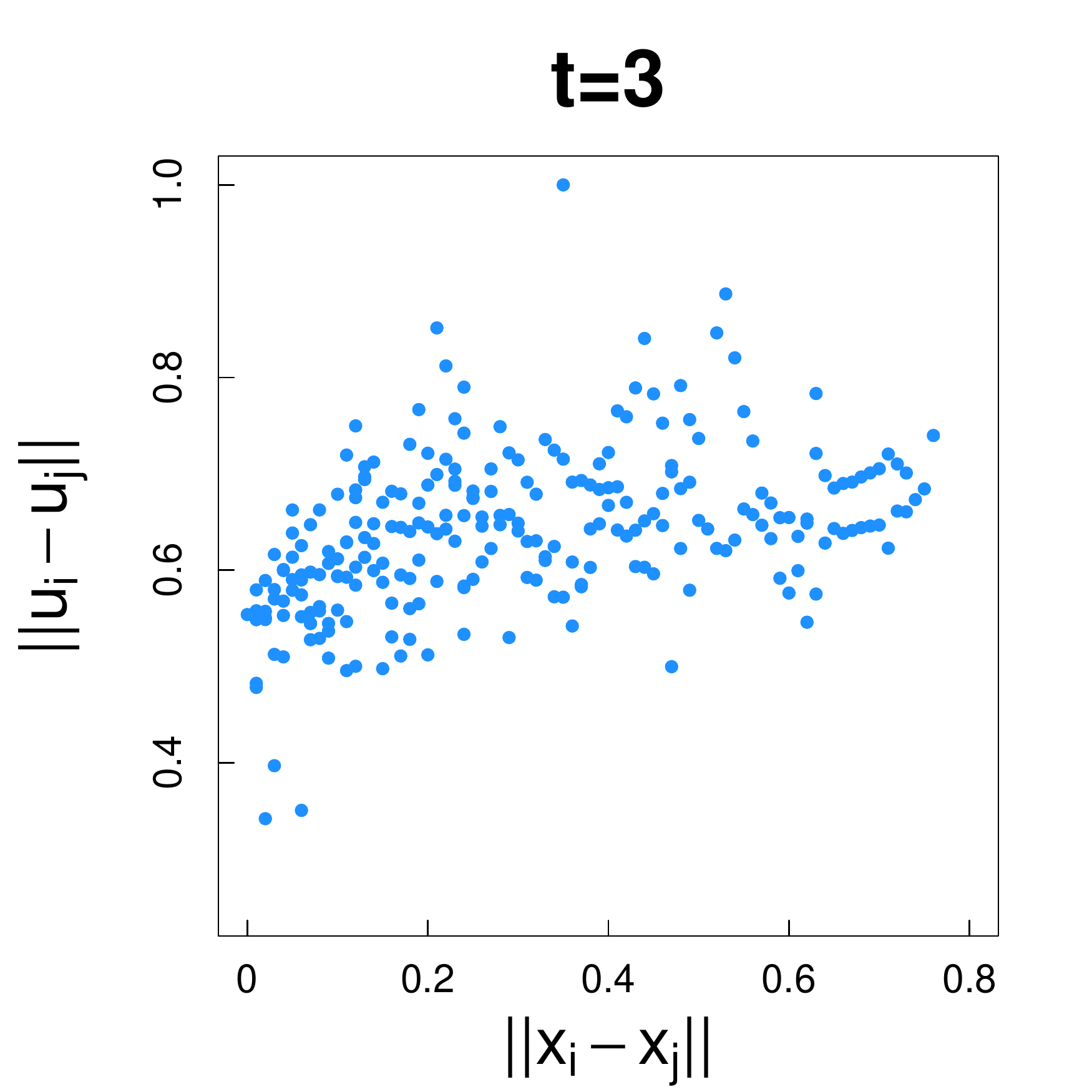} \\
		\centering (f) $\rho = 0.42$
	\end{minipage}
	\begin{minipage}[b]{0.20\textwidth}
		\includegraphics[width=\textwidth]{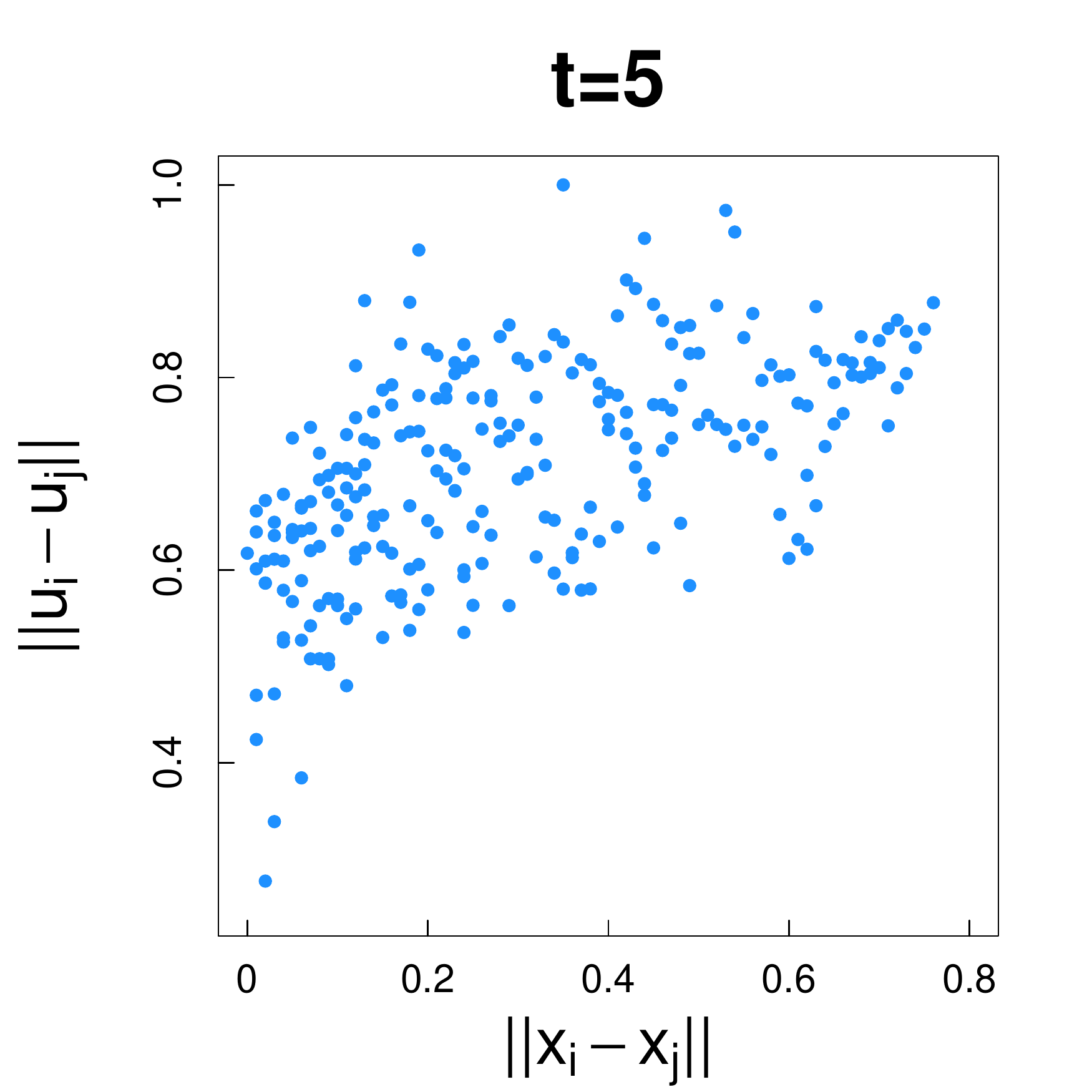} \\
		\centering (g) $\rho = 0.57$
	\end{minipage}
	\begin{minipage}[b]{0.20\textwidth}
		\includegraphics[width=\textwidth]{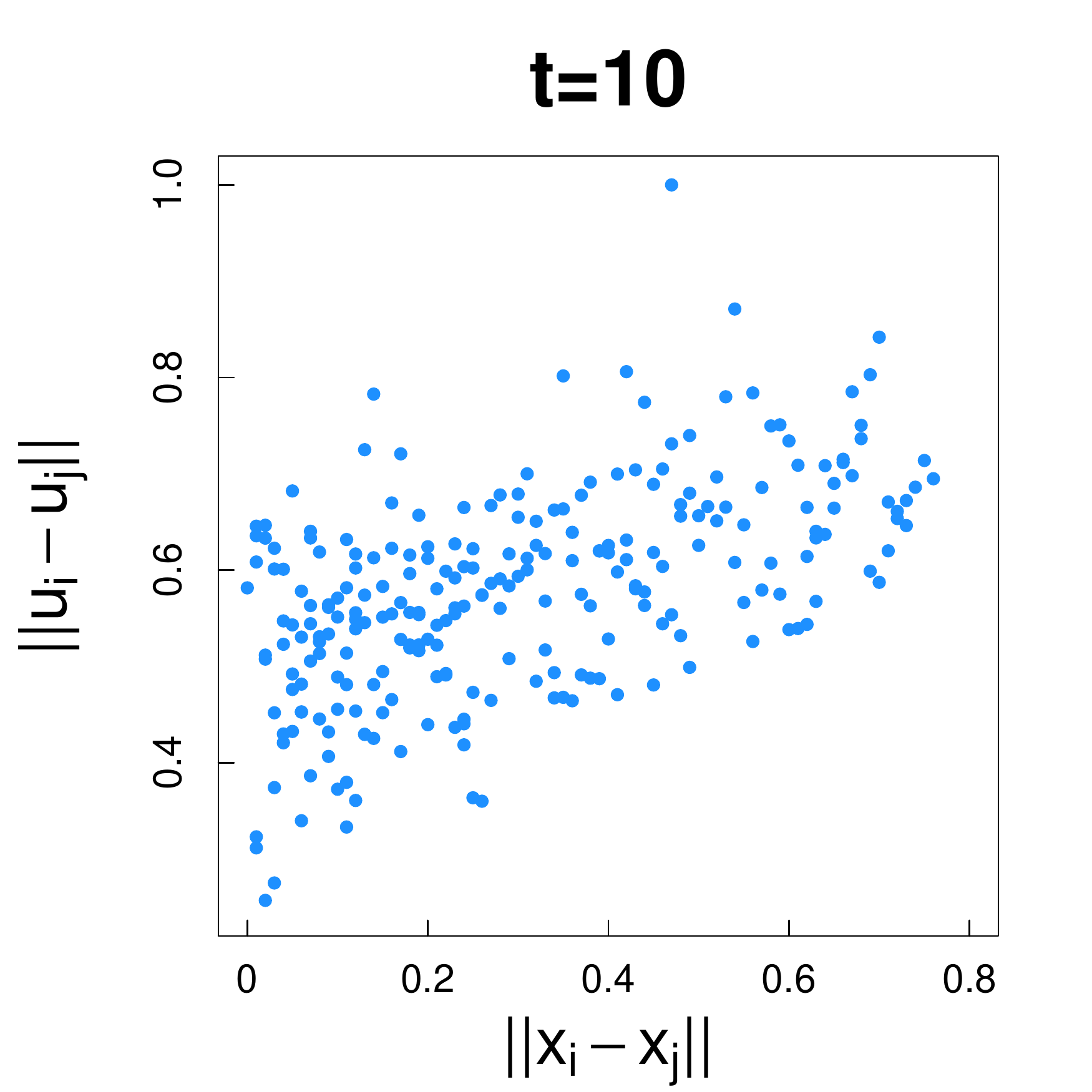} \\
		\centering (h) $\rho = 0.55$
	\end{minipage}
	\caption{Local correlation maps at different diffusion times. Figure~(c) presents the correlation map at optimal time $t^{*} = 5$ identified by diffusion multiscale graph correlation. Panel (e)-(h) show standardized Euclidean pairwise distance between $\{ U^{t}_{i}\}$ and $\{X_{i}\}$ for $t=1,3,5,10$, among which correlation between two distances is most evident at $t^{*}=5$ as well with highest correlation.}
	\label{fig:localcorr}
\end{figure}
Figure~\ref{fig:localcorr} (a)-(d) presents local distance correlation map $\dCorr^{kl}(\mc{U}, \mc{X})$ across diffusion times. These plots show that the optimal local correlation is detected at non-global neighborhood choice, i.e. $l^{*} \neq 68$ (the global maximum),
which imply a non-linear dependence between connectivity and position and an optimal $t^{*}=5$.
Figure~\ref{fig:localcorr} (e)-(h) illustrate the relationship between Euclidean distance in diffusion maps and nodal attributes at different diffusion times at $t=1,3,5,10$, which is again the most significant at $t=5$. 

\section{Discussion}

There are several potential follow-ups that would further advance the work. One example is more theoretical investigation into the smoothed maximum and dimension selection of $t^{\prime}$. Assuming $t^{\prime}$ is the true optimal diffusion time, it will be helpful to either identify a more systematic and reliable way to estimate $t^{\prime}$, or quantify variability in the estimated optimal $t^{*}$ by smoothed maximum. This would hopefully reduce computational burden instead of going over all possible diffusion times, e.g. $t=0,1,2, \ldots, 10$. Moreover, although we briefly discussed one example in Section~\ref{ssec:dis}, the impact of dimensional choice of $q$ is still obscure on the embedding quality. Finally, since one can apply diffusion map to any data and one can think of any affinity or kernel matrix as a graph, this method is actually applicable to more general testing scenarios beyond networks, which is another point of interest for further investigation.

\section*{Acknowledgement} 

The authors thank Dr.~Minh Tang, Dr.~Daniel Sussman, the editor and anonymous reviewers for their insightful suggestions to improve the paper.
This work was supported by 
the National Science Foundation,
and the Defense Advanced Research Projects Agency.
%the National Security Science and Engineering Faculty Fellowship (NSSEFF),
%
%the Johns Hopkins University Human Language Technology Center of Excellence (JHU HLT COE), 
%
%the Defense Advanced Research Projects Agency's (DARPA) SIMPLEX program through SPAWAR, the DARPA GRAPHS program, the DARPA D3M program through contract, and 
%
%the XDATA program of DARPA administered through Air Force Research Laboratory.
%

%\section*{Supplementary material}
%\label{SM}

%Supplementary material available at \Bka\ online includes proofs of the theoretical results, simulation settings of random dot product graphs. 

%\clearpage
\bibliographystyle{apalike}
\bibliography{reference}

%%%%%%%%%%%%%%%%%%%%%%%%%%%%%%%%%%%%%%%%%%%%%%
\clearpage
\appendix
\setcounter{figure}{0}
\renewcommand{\thefigure}{S\arabic{figure}}

\bigskip
\begin{center}
{\large\bf Supplementary Material}
\end{center}

\section{Proofs}
\label{sec:proof}
Unless mentioned otherwise, throughout the proof section we always omit the superscript $t$ for the diffusion map at a fixed $t$, i.e., we use $\mc{U}= \{U_i : i=1,2,\ldots,n\}$ instead of $\mc{U}^{t}= \{U_{i}^{t} : i=1,2,\ldots,n\}$ because most results hold for any $t$, similarly we use $\theta$ instead of $\theta^{t}$ whenever appropriate.

%%%%%%%%%%%%%%%%%%%%%%	
\begin{proof}[(Theorem 1)]
	By the \textit{de Finetti's Theorem}~\citep{diaconis1980finite,oneill2009,orbanz2015bayesian}, it suffices to prove that the diffusion map $\mc{U} = \{U_i: i=1,\ldots,n\}$ is always exchangeable in distribution, i.e., for any $n$ and all possible permutation $\sigma$, the permuted sequence $\mc{U}_{\sigma} = \{U_{\sigma(1)}, U_{\sigma(2)}, \ldots,U_{\sigma(n)}\}$ always distributes the same as the original sequence $\mc{U} = \{U_{1}, U_{2}, \ldots,U_{n}\}$.
	
	Transforming Equation 3 in the main manuscript into matrix notation yields
	\begin{align*}
	\mc{U} = \Lambda^{t}\Phi^{T},
	\end{align*}
	where $\mc{U}$ is the $q \times n$ matrix having $U_i$ as its $i^{\mbox{th}}$ column, $\Lambda=\mbox{diag} \{ \lambda_{1},\lambda_2,\ldots,\lambda_q \}$ is the diagonal matrix having selected eigenvalues of $\mc{L}$, $\Phi =[ \phi_1, \phi_2, \cdots, \phi_q ]$ consists of the corresponding eigenvectors, $\cdot^{t}$ denotes $t^{\mbox{th}}$ power, and $\cdot^{T}$ is the matrix transpose. It suffices to show that $\mc{U}$ and $\mc{U} \Pi$ are identically distributed for any permutation matrix $\Pi$ of size $n$.
	
	Given that the graph $\mc{G}$ is an induced subgraph of an infinitely exchangeable graph, it holds that $\mc{A}(\sigma(i),\sigma(j)) \stackrel{d}{=} \mc{A}(i,j)$, which further holds for the symmetric graph Laplacian $\mc{L}$:
	\begin{align*}
	\mc{L}(\sigma(i), \sigma(j)) &=  \mc{A}(\sigma(i), \sigma(j)) / \{ \sum\limits_{j} \mc{A}(\sigma(i), \sigma(j)) \sum\limits_{i} \mc{A}(\sigma(i), \sigma(j))\}^{1/2} \\
	&\stackrel{d}{=}  \mc{A}(i,j) / \{ \sum\limits_{j} \mc{A}(i,j) \sum\limits_{i} \mc{A}(i,j) \}^{1/2} \\
	&= \mc{L}(i,j).
	\end{align*}
	In matrix notation, $\Pi^{T} \mc{L} \Pi \stackrel{d}{=} L$ for any permutation matrix $\Pi$. 
	
	By eigen-decomposition, the first $q$ eigenvalues and the corresponding eigenvector of $\Pi^{T} \mc{L} \Pi$ are $\Lambda$ and $\Pi^{T} \Phi$, so it follows that at any $t$
	\begin{align*}
	& \Phi \stackrel{d}{=} \Pi^{T} \Phi\\
	\Leftrightarrow \quad
	& \mc{U} =\Lambda^{t} \Phi^{T}  \stackrel{d}{=} \Lambda^{t} \Phi^{T} \Pi = \mc{U} \Pi. 
	\end{align*}	
	Thus columns in $\mc{U}$ are exchangeable, i.e., the diffusion maps, $\{ U_{i} \in \mathbb{R}^{q} : i=1,2,\ldots, n \}$, are infinitely exchangeable. By the \textit{de Finetti's Theorem}, there exists an underlying variable $\theta$ distributed as the limiting empirical distribution, such that $U_{i} \mid \theta$ are asymptotically i.i.d. 
\end{proof}

%%%%%%%%%%%%%%%%%%%%%%%%
\begin{proof}[(Theorem 2)]
	
	We first state three lemmas:
	\begin{lemma}
		\label{lemma:aux1}
		Under the same assumptions of Theorem~2, for any finite time-step $t$, the underlying distribution of $U_{i}^{t}$ of the diffusion map is of finite second moment.
	\end{lemma}
	
	\begin{lemma}
		\label{lemma:aux2}
		The distance covariance of $(\mc{U}, \mc{X}) = \{  ( U_{i}, X_{i}  )  : i = 1, \ldots, n \}$ defined in Equation~5 in the paper satisfies
		\begin{align*}
		\dCov_{n}(\mc{U},\mc{X}) &=  \int_{\mathbb{R}^{q+p}} |\hat{g}_{\mc{U}, \mc{X}}(t,s)-\hat{g}_{\mc{U}}(t) \hat{g}_{\mc{X}} (s)|^{2} dw(t,s),
		\end{align*}
		where $w(t,s) \in \mathbb{R}^{q} \times \mathbb{R}^{p}$ is a nonnegative weight function that equals $(c_{q}c_{p}|t|_{q}^{1+q}|s|_{p}^{1+p})^{-1}$, $c_{q}$ is a nonnegative constant, $\hat{g}_{\cdot}$ is the empirical characteristic function of $\{(U_{i},X_{i}) : i=1,2,...,n\}$ or the marginals, e.g., $\hat{g}_{\mc{U},\mc{X}}(t,s)=\frac{1}{n}\sum_{i=1}^{n}\exp(\textbf{i} \left\langle t,U_{i} \right\rangle+\textbf{i} \left\langle s,X_{i} \right\rangle)$ with $\textbf{i}$ representing the imaginary unit.
	\end{lemma}
	
	% \jovo{the below notation is unclear}
	\begin{lemma}
		\label{lemma:aux3}
		Assume $\mc{U} = \{ U_{i} \sim F_U: i = 1,2, \ldots, n \}$ are conditional i.i.d. as $U \mid \theta$, and $\mc{X} = \{X_{i} \stackrel{i.i.d.}{\sim} F_{X}:~i = 1, 2, \ldots, n\}$, and all distributions are both of finite second moment. It follows that
		\begin{align*}
		\dCov_{n}(\mc{U}, \mc{X}) &\rightarrow \dCov(U,X) \mbox{ as } n \rightarrow \infty,
		\end{align*}
		where $\dCov(U,X) := \int_{\mathbb{R}^{q+p}} | g_{U,X}(t,s) - g_{U}(t) g_{X}(s) |^2 dw(t,s)$ is the population distance covariance, and $g_{\cdot}$ is the characteristic function, i.e., $g_{U,X}(t,s) = E(\exp\{\textbf{i} \left\langle t,U \right\rangle  +\textbf{i} \left\langle  s, X\right\rangle \})$.
	\end{lemma}
	
	By Theorem~1, the diffusion maps $U_{i}$ are asymptotically i.i.d. conditioned on $\theta$, whose finite moment is guaranteed by Lemma~\ref{lemma:aux1}. The nodal attributes $X_i$ are i.i.d. as $F_{X}$ of finite second moment as assumed in (C2). Therefore a direct application of Lemma~\ref{lemma:aux2} and Lemma~\ref{lemma:aux3} yields that
	\begin{align*}
	\mbox{\dCov}_{n}(\mc{U} ,\mc{X}) &\rightarrow \int_{\mathbb{R}^{q+p}} | g_{U,X}(t,s) - g_{U}(t) g_{X}(s) |^2 dw(t,s),
	\end{align*}
	which equals $0$ if and only if $U$ is independent of $X$. As distance correlation is just a normalized version of distance covariance, it further leads to
	\begin{align}
	\label{eq:main1}
	\mbox{\dCorr}_{n}(\mc{U},\mc{X}) &\rightarrow c \geq 0,
	\end{align}
	for which the equality holds if and only if $F_{UX}=F_{U}F_{X}$. By \cite{shen2017mgc}, Equation~\ref{eq:main1} also holds for \MGC~when it holds for \dCorr. %Therefore, \MGC is consistent for testing dependence between the diffusion maps $U$ and the nodal attributes $X$.
\end{proof}
%%%%%%%%%%%%%%%%%%%%%%%%%%%%%%%%%%%%%%%%%%%%%
\begin{proof}[(Lemma~\ref{lemma:aux1})]
	To prove that $U$ is of finite second moment, it suffices to show that $\|U_{i}\|_{2}$ is always bounded for all $i \in [1,n]$. By Equation~3, we have 
	\begin{align*}
	\| U_i \|_{2}^{2}  &= \sum_{j=1}^{q} \lambda^{2t}_{j}\phi_{j}^{2}(i) \\ 
	& \leq \sum_{j=1}^{q} \lambda^{2t}_{j} \\
	& \leq q,
	\end{align*}
	where the second line follows by noting $\phi_{j}(i) \in [-1,1]$ (the eigenvector $\phi_{j}$ is always of unit norm), and the third line follows by observing that $| \lambda_{j}| \leq \| L \|_{\infty}= 1$. Therefore, all of $U_{i}$ are bounded in $\ell_{2}$ norm as $n \rightarrow \infty$, so the underlying variable $U$ must be of finite second moment for any finite $t$.
\end{proof}

\begin{proof}[(Lemma~\ref{lemma:aux2})] 
	This lemma is a direct application of Theorem 1 in \cite{szekely2007measuring}, which holds without any assumption on $(\mc{U} ,\mc{X}) =  \{(U_{i},X_{i}) :  i=1,2,...,n\}$, e.g., it holds without assuming exchangeability, nor identically distributed, nor finite moment.
\end{proof}

\begin{proof}[(Lemma~\ref{lemma:aux3})] 
	This lemma is equivalent to Theorem 2 in \cite{szekely2007measuring}, except the i.i.d. assumption is replaced by exchangeable assumption, i.e., the original set-up needs $(\mc{U}, \mc{X}) = \{(U_{i},X_{i}) : i = 1,2, \ldots , n \}$ to be independently identically distributed as $F_{UX}$ with finite second moment; whereas the diffusion map $\{  U_{i} : i = 1,2, \ldots, n \}$ is asymptotically conditional i.i.d. with finite second moment.
	
	Note that $\hat{g}_{\mc{U}, \mc{X}}(t,s)=E(\hat{g}_{\mc{U}, \mc{X} }(t,s) \mid \theta)$, and each term in $\hat{g}_{\mc{U}, \mc{X}}(t,s) \mid \theta$ is asymptotically i.i.d. of each other. Thus
	\begin{align*}
	\displaystyle\int{|\hat{g}_{\mc{U},\mc{X}} (t,s)-\hat{g}_{\mc{U}}(t)\hat{g}_{\mc{X}}(s)|^{2}}dw &=
	E(\displaystyle\int{|\hat{g}_{\mc{U},\mc{X}} (t,s)-\hat{g}_{\mc{U}}(t)\hat{g}_{\mc{X}}(s)|^{2}}dw \mid \theta) \\
	& \rightarrow E( \int{|g_{U,X}(t,s)-g_{U}(t)g_{X}(s)|^{2}}dw \mid \theta) \\
	& = \int{|g_{U,X}(t,s)-g_{U}(t)g_{X}(s)|^{2}}dw,
	%\label{eq:SLLN}
	\end{align*}
	where the convergence in the second step follows from Theorem 2 in \cite{szekely2007measuring} on the i.i.d. case.
	%All major steps being essentially the same for either i.i.d. random variables or infinitely exchangeable random variables 
	%\citep{InoueTaylor2006}
	%, under the finite moment assumption we still have the large number for V-statistics 
	%\citep{KoroljukBook}
	%, i.e., 
	%\begin{align*}
	%\displaystyle\int_{D(\delta)}{|\hat{g}_{\mathbf{u},\mathbf{x}} (t,s)-\hat{g}_{\mathbf{u}}(t)\hat{g}_{\mathbf{x}}(s)|^{2}}dw &\stackrel{n \rightarrow \infty}{\longrightarrow} 
	%\displaystyle\int_{D(\delta)}{|g_{\mathbf{u},\mathbf{x}}(t,s)-g_{\mathbf{u}}(t)g_{\mathbf{x}}(s)|^{2}}dw,
	%\label{eq:SLLN}
	%\end{align*}
	%where $D(\delta)=\{(t,s):\delta \leq |t|_{q} \leq 1/\delta,\delta \leq |s|_{p} \leq 1/\delta\}$. %and the nonnegative weight function $w(t,s)$ is chosen by \cite{szekely2007measuring}. 
\end{proof}

\begin{proof}[(Theorem~3)] 
	From Theorem~2, it holds that 
	\begin{align}
	\label{eq:mgczero}
	\MGC_{n}(\mc{U}^{t},\mc{X}) &\rightarrow c \geq 0
	\end{align}
	for each $t$, with equality if and only if independence. The \DMGC~algorithm enforces that 
	\begin{align*}
	\max\{\MGC_{n}(\mc{U}^{t},\mc{X}), t=0,1,\ldots,10\} &\geq \MGC_{n}^{*}(\{\mc{U}^{t}\},\mc{X}) \geq \MGC_{n}(\mc{U}^{t=3},\mc{X}),
	\end{align*}
	thus Equation~\ref{eq:mgczero} also holds when $\MGC(\mc{U}^{t},\mc{X})$ is replaced by $\MGC^{*}(\{\mc{U}^{t}\},\mc{X})$.
	
	To show that the test is valid and consistent, it suffices to show that with probability approaching $1$, $\MGC_{n}(\mc{U},\mc{X}_{\sigma}) \rightarrow 0$. This holds when $(U_{i}, X_{i}) \overset{i.i.d.}{\sim} F_{UX}$: the proof in supplementary of \cite{shen2017mgc} shows that the percentage of partial derangement of finite sample size converges to $1$ among all random permutations, such that with probability converging to $1$ a permutation test breaks dependency.
	
	For exchangeable $\{U_{i}\}$ here, we instead have $(U_{i}, X_{i}) \mid \theta \overset{i.i.d.}{\sim} F_{UX \mid \theta}$ asymptotically. The distribution of $\theta$ is the limiting empirical distribution of $\{U_{i}\}$, which is either asymptotically independent of all $X_i$ or dependent only on finite number of $X_i$. Thus $U_{i}$ is asymptotically conditionally independent with $X_{\sigma{(i)}}$ with probability converging to $1$, and we have
	\begin{align*}
	\MGC_{n}(\mc{U},\mc{X}_{\sigma})
	=E(\MGC_{n}(\mc{U},\mc{X}_{\sigma}) \mid \theta)
	\rightarrow 0
	\end{align*}
	Moreover, when the transformation from $\mc{A}$ to $\mc{U}^{t}$ is injective, we have
	\begin{align*}
	&\mbox{ $A$ is independent of $X$} \\
	\Leftrightarrow & \mbox{ $U^{t}$ is independent of $X$ for all $t$} \\
	\Leftrightarrow & \mbox{ $\MGC_{n}(\mc{U}^{t},\mc{X})$ is asymptotically $0$,}
	\end{align*}
	where the second line follows from injective transformation, and the third line follows from Theorem~1 and Theorem~2. Thus \DMGC~is consistent between $A$ and $X$.
	
	Note that without the injective condition, the reverse direction of the second line may not always hold, i.e., when the diffusion maps are independent from the nodal attributes, the adjacency matrix may be still dependent with the nodal attributes. In that case, \DMGC~is still valid but the dependency may not be detected by \DMGC.
\end{proof}

\begin{proof}[(Corollary~1)] 
	%(1) All the proofs remain the same regardless of the choice of $t$ and $q$, as long as they are finite positive integers.
	%(1) When the nodal attribute is replaced by a second graph with the same node set as the first graph, we simply compute the diffusion map of the second graph by the same procedure as for the first graph. %The resulting diffusion map of the second graph satisfies both Theorem~{lemma:LSE} and Theorem~{theorem:convergence}.
	
	(1) Changing the test statistic only affects Theorem~2. Both \dCorr~and~\MGC~satisfy Theorem~2 directly, while \HHG~is also a statistic that is $0$ if and only if independence~\citep{HellerGorfine2013}.
	
	(2) %If the adjacency matrix $A$ is symmetric and positive semi-definite, so is the normalized Laplacian matrix $L$. In particular, they share the same eigenvalues.
	When $\mc{A}$ is symmetric and binary, the transformation from $\mc{A}$ to $\mc{L}$ is injective, i.e., two different $\mc{A}$ always produce two different $\mc{L}$. Then for each unique $\mc{L}$, the eigen-decomposition is always unique such that $\mc{L}$ to $\mc{U}^{t=1}$ is injective, provided that the dimension choice is made correct at $q$.
\end{proof}

%%%%%%%%%%%%%%%%%%%%%%%%%%%%%%%
\begin{proof}[(Corollary~2)] 
	From Proposition~1 and~2,~$\mc{U}^{t=1}$ is asymptotically equivalent to the latent positions $\mc{W}$ up-to a bijection. Moreover, under random dot product graph, if two different adjacency matrices yield the same $\mc{U}^{t=1}$, they must asymptotically equal the same latent positions and asymptotically the same adjacency matrix (i.e., the difference in Frobenius norm converges to $0$). Therefore injective holds asymptotically, and Theorem~3 applies.
\end{proof}

\newpage
\section{Additional Simulation}
\label{sec:addsim}

Here we investigate the performance of the test statistics under the violation of positive semi-definite link function related to condition (C3) in the main paper. Under the non-random dot product graph, we generate following two-block stochastic block model for $n=100$:
\begin{equation}
\begin{split}
\label{eq:nonposi}
& Z_{i}  \overset{i.i.d.}{\sim} \mc{B}(0.5), \\
& \mc{A}(i,j) \mid Z_{i}, Z_{j}  \sim Bernoulli \left\{ (0.5-\epsilon) \mathbb{I}\left( |Z_{i} - Z_{i}| = 0 \right) + 0.3 \mathbb{I}\left( |Z_{i} - Z_{j}| \neq 0 \right) \right\}, \\ 
& X_{i} \mid Z_{i}  \sim \mc{B} \left(  Z_{i} / 3 \right),
\end{split}
\end{equation}
where $Z_{i}$ represents the block membership, $X_{i}$ is the noisy membership from $Z_{i}$, and we test independence between $\mc{A}$ and $Z$. When $\epsilon = 0.2$, $\mc{A}$ and $Z$ are actually independent.
When $\epsilon > 0.2$, the above model yields a non-positive semi-definite graph. As $\epsilon$ increases from $0.2$, the dependency signal gets stronger.
%Above Equation~\ref{eq:nonposi} results non-positive semi-definite graph when $\epsilon > 0.2$, and beyond $\epsilon > 0.2$ increasing $\epsilon$ implies larger dependency.

\begin{figure}[H]
	\centering		\includegraphics[width=0.7\textwidth]{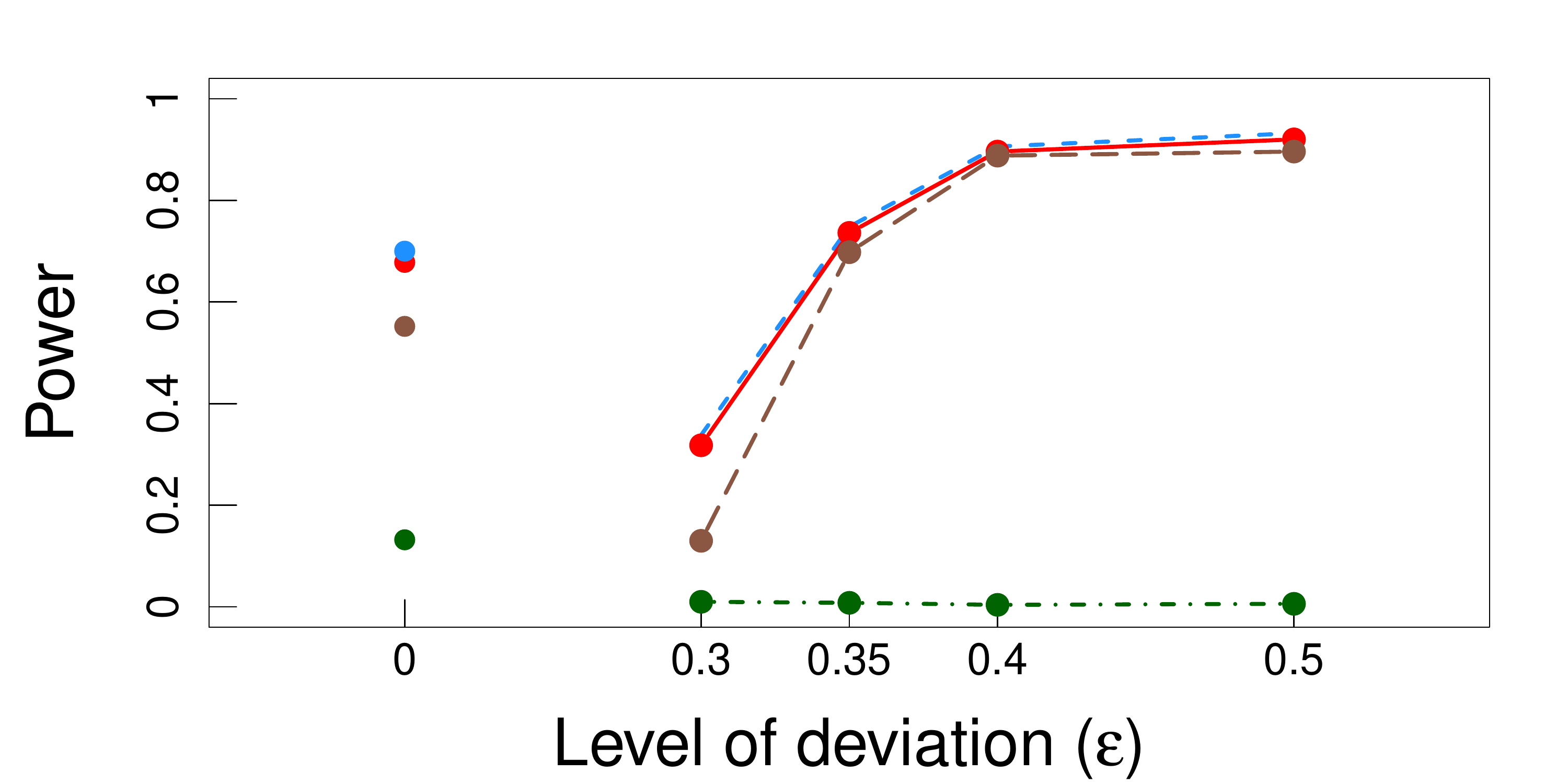}	
	\caption{\label{fig:bmatrix}
		Diffusion map-based dependency measures are robust against non-positive semi-definite link function (see equation~\ref{eq:nonposi} for details). All of diffusion~\MGC~(red solid), diffusion~\dCorr~(yellow brown dashes), and diffusion~\HHG~(blue small dashes) have better power than the
		\FH~test (green dot-dashes). %In fact power of distance-based test statistics increases as discrepancy in edge probability between two blocks increases with increasing $\epsilon$ regardless deviation from positive semi-definiteness. All the testing powers are derived from $m=500$ random replicates for which each $p$-value is from $r=500$ permutations.
	}
\end{figure}
% total rewrite of caption
% Figure~\ref{fig:bmatrix} shows that distance-based methods, i.e., \MGC, \dCorr, and \HHG, all preserve testing power under $\epsilon > 0.2$; while the Fosdick-Hoff likelihood-based test does not.

\newpage
\section{Random Dot Product Graph Simulations}
\label{sec:RDPG_model}

\begin{figure}[H]
	\centering
	\includegraphics[width=0.7\textwidth]{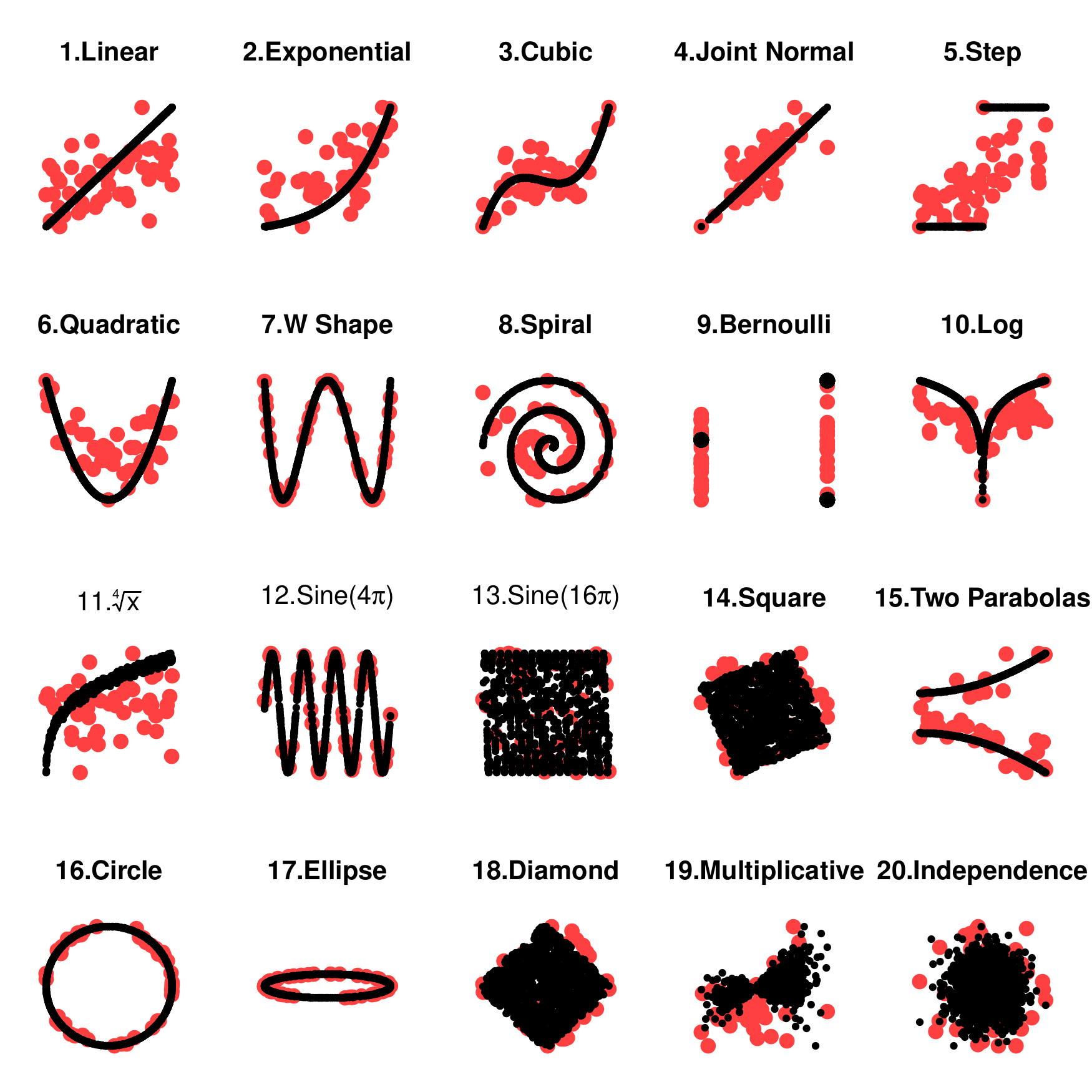}	
	\caption{Illustrations of randomly generated $n=50$ points of $\{ (W_{i}, X_{i}) : i=1,2,\ldots, 50 \}$ (red dots) along with their population version without noise (black dots).}
	\label{fig:RDPG_setting}
\end{figure} 
%\figurebox{20pc}{25pc}{}[]

For the $20$ simulations under random dot product graph, we describe the generating distribution ($\tilde{W}_{i}, \tilde{X}_{i}) \overset{i.i.d.}{\sim} F_{\tilde{W}, \tilde{X}}$ under each scenario. %Except for 17.Ellipse, both of $w_{i}$ and $x_{i}$ are squeezed into [0,1] as described in Section~{ssec:RDPG}. In case of 17.Ellipse, $x_{i} = \tilde{x}_{i} / (\max\{ \tilde{w}_{j} : j=1,2,\ldots, n \}  - \min \{ \tilde{w}_{j} : j=1,2,\ldots, n \} ) + 0.5$ in order to preserve the shape. 
They are based on \cite{shen2016discovering, shen2017mgc}, and visualization for the sample observations of $\{ (W_{i}, X_{i}) : i=1,2,\ldots, n=50 \}$ is shown in Fig.\ref{fig:RDPG_setting}. Notation-wise, $\mc{N}(\mu, \sigma)$ denotes the normal distribution with mean $\mu$ and standard deviation $\sigma$, $\mc{U}[a,b]$ denotes the uniform distribution from $a$ to $b$, $\mc{B}(p)$ denotes the Bernoulli distribution with probability $p$, and $\epsilon_{i}$ denotes white noise.

1. Linear
\begin{align*}
\tilde{W}_{i} & \sim \mc{U}[0,1],~\epsilon_{i} \sim \mc{N}(0,0.5), \\
\tilde{X}_{i} & =  \tilde{W}_{i} + \epsilon_{i}.
\end{align*}

2. Exponential
\begin{align*}
\tilde{W}_{i} & \sim \mc{U}[0,3],~\epsilon_{i} \sim \mc{N}(0,5), \\
\tilde{X}_{i} & =  \exp(\tilde{W}_{i}) + \epsilon_{i}.
\end{align*}

3. Cubic
\begin{align*}
\tilde{W}_{i} & \sim \mc{U}[0,1],~\epsilon_{i} \sim \mc{N}(0,0.5), \\
\tilde{X}_{i} & =  20(\tilde{W}_{i} - 0.5)^3 + 2 (\tilde{W}_{i} - 0.5)^2 - (\tilde{W}_{i} - 0.5) + \epsilon_{i}.
\end{align*}

4. Joint Normal
\begin{align*}
(\tilde{W}_{i}, \tilde{X}_{i}) & \sim \mc{N} \left( \begin{bmatrix} 0 \\ 0 \end{bmatrix}, \begin{bmatrix} 0.7 & 0.5 \\ 0.5 & 0.7 \end{bmatrix}  \right). \\
\end{align*}

5. Step Function
\begin{align*}
\tilde{W}_{i} & \sim \mc{U}[-1,1],~\epsilon_{i} \sim \mc{N}(0,0.5), \\
\tilde{X}_{i} & = \mathbb{I}(\tilde{W}_{i} > 0 ) + \epsilon_{i}.
\end{align*}

6. Quadratic
\begin{align*}
\tilde{W}_{i} & \sim \mc{U}[-1,1],~\epsilon_{i} \sim \mc{N}(0,0.3), \\
\tilde{X}_{i} & = \tilde{W}^2_{i} + \epsilon_{i}.
\end{align*}

7. W Shape
\begin{align*}
\tilde{W}_{i} & \sim \mc{U}[-1,1] \\
\tilde{X}_{i} & = 4( \tilde{W}^2_{i} - 0.5 )^2 
\end{align*}

8. Spiral
\begin{align*}
Z_{i} & \sim \mc{U}[0,5],~\epsilon_{i} \sim \mc{N}(0,0.1), \\
\tilde{W}_{i} & = Z_{i} \cos ( Z_{i} \pi), \\
\tilde{X}_{i} & =  Z_{i} \sin (Z_{i} \pi) + \epsilon_{i}.
\end{align*}

9. Bernoulli
\begin{align*}
\tilde{W}_{i} & \sim \mc{B}(0.5),~\epsilon_{i} \sim \mc{N}(0,1), \\
\tilde{X}_{i} & =  (2 \mc{B}(0.5) - 1)\tilde{W}_{i} + \epsilon_{i}.
\end{align*}

10. Logarithm
\begin{align*}
\tilde{W}_{i} &\sim \mc{U}[-1, 1],~\epsilon_{i} \sim \mc{N}(0,5), \\
\tilde{X}_{i} & = 5 \log_{2} (|\tilde{W}_{i}|) + \epsilon_{i}.
\end{align*}

11. Fourth Root
\begin{align*}
\tilde{W}_{i} & \sim \mc{U}[0, 1],~\epsilon_{i} \sim \mc{N}(0,0.5), \\
\tilde{X}_{i} & = {(|\tilde{W}_{i} + \epsilon_{i}|)}^{1/4}.
\end{align*}

12. Sine Period 4$\pi$
\begin{align*}
\tilde{W}_{i} & \sim \mc{U}[-1, 1],~\epsilon_{i} \sim \mc{N}(0,0.01), \\
\tilde{X}_{i} & = \sin ( 4 \tilde{W}_{i} \pi ) + \epsilon_{i}.
\end{align*}

13. Sine Period 16$\pi$
\begin{align*}
\tilde{W}_{i} & \sim \mc{U}[-1, 1],~\epsilon_{i} \sim \mc{N}(0,0.01), \\
\tilde{X}_{i} & = \sin ( 16 \tilde{W}_{i} \pi ) + \epsilon_{i}.
\end{align*}

14. Square
\begin{align*}
U_{i1} & \sim \mc{U}[-1, 1], ~ u_{i2} \sim \mc{U}[-1, 1], \\ 
\tilde{W}_{i} & =  U_{i1} \cos(-\pi/8) + U_{i2} \sin(-\pi/8),  \\
\tilde{X}_{i} & =  -U_{i1} \sin(-\pi / 8) + U_{i2} \cos(-\pi/8).
\end{align*}

15.Two Parabolas
\begin{align*}
\tilde{Z}_{i} &\sim \mc{B}(0.3),~\epsilon_{i} \sim \mc{N}(0.5, 0.3),\\
\tilde{W}_{i} & \sim \mc{U}[0, 1], \\
\tilde{X}_{i} & = (\tilde{W}^{2}_{i} + \epsilon_{i} ) (\tilde{Z}_{i} - 0.5).
\end{align*}

16. Circle
\begin{align*}
U_{i} & \sim \mc{U}[-1,1],~\epsilon_{i} \sim \mc{N}(0, 0.05),\\
\tilde{W}_{i} & = \cos( U_{i} \pi), \\
\tilde{X}_{i} & = \sin (U_{i} \pi) + \epsilon_{i}.
\end{align*}

17. Ellipse
\begin{align*}
U_{i} & \sim \mc{U}[-1,1], \\
\tilde{W}_{i} & = 5\cos( U_{i} \pi),  \\
\tilde{X}_{i} & = \sin (U_{i} \pi) .
\end{align*}

18. Diamond
\begin{align*}
U_{i1} & \sim \mc{U}[-1, 1], ~ U_{i2} \sim \mc{U}[-1, 1], \\
\tilde{W}_{i} & =  U_{i1} \cos(-\pi/4) + U_{i2} \sin(-\pi/4),  \\
\tilde{X}_{i} & =  -U_{i1} \sin(-\pi / 4) + U_{i2} \cos(-\pi/4).
\end{align*}

19. Multiplicative Noise
\begin{align*}
\tilde{W}_{i} & \sim \mc{N}(0.5, 1),~\epsilon_{i} \sim \mc{N}(0.5, 1),\\ 
\tilde{X}_{i} & = \tilde{W}_{i} \cdot \epsilon_{i}
\end{align*}

20. Independence
\begin{align*}
\tilde{W}_{i} & \sim \mc{N}(0,1) \\
\tilde{X}_{i} & \sim \mc{U}(0,1) \\
\end{align*}

\end{document}